\input harvmac
\input epsf
\noblackbox
\def\a{\alpha}
\def\p{{\partial}}

\def\ai{\alpha_{in}}
\def\o{\omega}

\lref\DiFrancescoSS{ P.~Di Francesco and D.~Kutasov,
Phys.\ Lett.\ B {\bf 261}, 385 (1991).
} \lref\FriessTQ{ J.~J.~Friess and H.~Verlinde, ``Hawking effect
in 2-D string theory,'' arXiv:hep-th/0411100.
}
\lref\aj{
J.~L.~Karczmarek and A.~Strominger,
``Closed string tachyon condensation at c = 1,''
JHEP {\bf 0405}, 062 (2004)
[arXiv:hep-th/0403169].
}

\lref\finn{
S.~R.~Das, J.~L.~Davis, F.~Larsen and P.~Mukhopadhyay,
``Particle production in matrix cosmology,''
Phys.\ Rev.\ D {\bf 70}, 044017 (2004)
[arXiv:hep-th/0403275].}

\lref\aaj{J.~L.~Karczmarek, A.~Maloney and A.~Strominger, ``Hartle-Hawking vacuum for c = 1 tachyon condensation,''
arXiv:hep-th/0405092.}

\lref\dvv{ R.~Dijkgraaf, H.~Verlinde and E.~Verlinde, ``String
propagation in a black hole geometry,'' Nucl.\ Phys.\ B {\bf 371},
269 (1992).}
\lref\MandalTZ{
G.~Mandal, A.~M.~Sengupta and S.~R.~Wadia,
``Classical solutions of two-dimensional string theory,''
Mod.\ Phys.\ Lett.\ A {\bf 6}, 1685 (1991).
}
\lref\GukovYP{
S.~Gukov, T.~Takayanagi and N.~Toumbas,
``Flux backgrounds in 2D string theory,''
JHEP {\bf 0403}, 017 (2004)
[arXiv:hep-th/0312208].
}
\lref\DiFrancescoUD{ P.~Di Francesco and D.~Kutasov,
Nucl.\ Phys.\ B {\bf 375}, 119 (1992) [arXiv:hep-th/9109005].
}
\lref\Martinec{ E.~Martinec and K.~Okuyama, ``Scattered results in
2D string theory,'' arXiv:hep-th/0407136.
}
\lref\polchcons{
J.~Polchinski,
 ``On the nonperturbative consistency of d = 2 string theory,''
Phys.\ Rev.\ Lett.\  {\bf 74}, 638 (1995)
[arXiv:hep-th/9409168].
}

\lref\GinspargIS{ P.~H.~Ginsparg and G.~W.~Moore,
arXiv:hep-th/9304011.
}

\lref\DasHW{
S.~R.~Das, J.~L.~Davis, F.~Larsen and P.~Mukhopadhyay,
``Particle Production in Matrix Cosmology,''
arXiv:hep-th/0403275.
}

\lref\SchomerusVV{ V.~Schomerus, ``Rolling tachyons from Liouville
theory,'' arXiv:hep-th/0306026.
}

\lref\ftt{M.~Headrick, S.~Minwalla and T.~Takayanagi, ``Closed
string tachyon condensation: an overview'', to be published.}
\lref\MooreZV{ G.~W.~Moore, M.~R.~Plesser and S.~Ramgoolam,
``Exact S matrix for 2-D string theory,'' Nucl.\ Phys.\ B {\bf
377}, 143 (1992) [arXiv:hep-th/9111035].
}
\lref\MooreSF{ G.~W.~Moore, ``Double scaled field theory at c = 1,''
Nucl.\ Phys.\ B {\bf 368}, 557 (1992).
}

\lref\SeibergBJ{ N.~Seiberg and S.~H.~Shenker, ``A Note on
background (in)dependence,'' Phys.\ Rev.\ D {\bf 45}, 4581 (1992)
[arXiv:hep-th/9201017].
}
\lref\Francesco{
P.~Di Francesco and D.~Kutasov,
 ``World sheet and space-time physics in two-dimensional (Super)string
 theory,''
Nucl.\ Phys.\ B {\bf 375}, 119 (1992)
[arXiv:hep-th/9109005].
}
\lref\dgik{
D.~J.~Gross and I.~R.~Klebanov,
 ``S = 1 for c = 1,''
Nucl.\ Phys.\ B {\bf 359}, 3 (1991).
}

\lref\WittenYR{ E.~Witten, ``On string theory and black holes,''
Phys.\ Rev.\ D {\bf 44}, 314 (1991).
}
\lref\TakayanagiSM{ T.~Takayanagi and N.~Toumbas, ``A matrix model
dual of type 0B string theory in two dimensions,'' JHEP {\bf
0307}, 064 (2003) [arXiv:hep-th/0307083].
}

\lref\DaCunhaFM{ B.~C.~Da Cunha and E.~J.~Martinec,
Phys.\ Rev.\ D {\bf 68}, 063502 (2003) [arXiv:hep-th/0303087].
}

\lref\GrossAY{ D.~J.~Gross and N.~Miljkovic, ``A Nonperturbative
Solution of $D = 1$ String Theory,'' Phys.\ Lett.\ B {\bf 238},
217 (1990);
}
%
\lref\BrezinSS{ E.~Brezin, V.~A.~Kazakov and A.~B.~Zamolodchikov,
``Scaling Violation in a Field Theory of Closed Strings in One
Physical Dimension,'' Nucl.\ Phys.\ B {\bf 338}, 673 (1990);
}

\lref\sshe{S. Shenker, unpublished.}

\lref\MukhopadhyayFF{
P.~Mukhopadhyay,
JHEP {\bf 0408}, 032 (2004)
[arXiv:hep-th/0406029].
}

\lref\GinspargAS{ P.~Ginsparg and J.~Zinn-Justin, ``2-D Gravity +
1-D Matter,'' Phys.\ Lett.\ B {\bf 240}, 333 (1990).
}
\lref\DasKA{ S.~R.~Das and A.~Jevicki, ``String Field Theory And
Physical Interpretation Of D = 1 Strings,'' Mod.\ Phys.\ Lett.\ A
{\bf 5}, 1639 (1990).
}
\lref\DouglasUP{ M.~R.~Douglas, I.~R.~Klebanov, D.~Kutasov,
J.~Maldacena, E.~Martinec and N.~Seiberg, ``A new hat for the c =
1 matrix model,'' arXiv:hep-th/0307195.
}
\lref\StromingerFN{ A.~Strominger and T.~Takayanagi, ``Correlators
in timelike bulk Liouville theory,'' arXiv:hep-th/0303221.
}

\lref\kkk{ V.~Kazakov, I.~K.~Kostov and D.~Kutasov, `A matrix
model for the two-dimensional black hole,'' Nucl.\ Phys.\ B {\bf
622}, 141 (2002) [arXiv:hep-th/0101011].
}
\lref\GrossUB{ D.~J.~Gross and I.~R.~Klebanov, ``One-Dimensional
String Theory On A Circle,'' Nucl.\ Phys.\ B {\bf 344}, 475
(1990).
}
\lref\sen{ A.~Sen, ``Rolling tachyon,'' JHEP {\bf 0204}, 048
(2002) [arXiv:hep-th/0203211].
}
\lref\mgas{ M.~Gutperle and A.~Strominger, ``Spacelike branes,''
JHEP {\bf 0204}, 018 (2002) [arXiv:hep-th/0202210].
}
\lref\MinicRK{ D.~Minic, J.~Polchinski and Z.~Yang, ``Translation
Invariant Backgrounds In (1+1)-Dimensional String Theory,'' Nucl.\
Phys.\ B {\bf 369}, 324 (1992).
}
\lref\MartinecKA{ E.~J.~Martinec, ``The annular report on
non-critical string theory,'' arXiv:hep-th/0305148.
}
\lref\PolchinskiMB{ J.~Polchinski, ``What is string theory?,''
arXiv:hep-th/9411028.
}

\lref\kms{ I.~R.~Klebanov, J.~Maldacena and N.~Seiberg, ``D-brane
decay in two-dimensional string theory,'' JHEP {\bf 0307}, 045
(2003) [arXiv:hep-th/0305159].
}
\lref\GinspargIS{ P.~Ginsparg and G.~W.~Moore, ``Lectures On 2-D
Gravity And 2-D String Theory,'' arXiv:hep-th/9304011.
}
\lref\hv{ J.~McGreevy and H.~Verlinde, ``Strings from tachyons:
The c = 1 matrix reloaded,'' arXiv:hep-th/0304224.
}
\lref\msy{ A.~Maloney, A.~Strominger and X.~Yin, ``S-brane
thermodynamics,'' arXiv:hep-th/0302146.
}

\lref\bd{ N.~D.~Birrell and P.~C.~W.~Davies, ``Quantum Fields In
Curved Space,'' Cambridge Univ. Pr. (1982).
}

\lref\PolchinskiUQ{ J.~Polchinski, ``Classical Limit Of
(1+1)-Dimensional String Theory,'' Nucl.\ Phys.\ B {\bf 362}, 125
(1991).
} \lref\KlebanovQA{ I.~R.~Klebanov, ``String theory in
two-dimensions,'' arXiv:hep-th/9108019.
}

\lref\NatsuumeSP{ M.~Natsuume and J.~Polchinski, ``Gravitational
Scattering In The C = 1 Matrix Model,'' Nucl.\ Phys.\ B {\bf 424},
137 (1994) [arXiv:hep-th/9402156].
}

\lref\KarczmarekXM{ J.~L.~Karczmarek, H.~Liu, J.~Maldacena and
A.~Strominger, ``UV finite brane decay,'' arXiv:hep-th/0306132.
}

\lref\AlexandrovCM{ S.~Alexandrov and V.~Kazakov, ``Correlators in
2D string theory with vortex condensation,'' Nucl.\ Phys.\ B {\bf
610}, 77 (2001) [arXiv:hep-th/0104094].
}

\lref\AlexandrovFH{ S.~Y.~Alexandrov, V.~A.~Kazakov and
I.~K.~Kostov, ``Time-dependent backgrounds of 2D string theory,''
Nucl.\ Phys.\ B {\bf 640}, 119 (2002) [arXiv:hep-th/0205079].
}

\lref\AlexandrovPZ{ S.~Y.~Alexandrov and V.~A.~Kazakov,
``Thermodynamics of 2D string theory,'' JHEP {\bf 0301}, 078
(2003) [arXiv:hep-th/0210251].
}

\lref\AlexandrovQK{ S.~Y.~Alexandrov, V.~A.~Kazakov and
I.~K.~Kostov, ``2D string theory as normal matrix model,'' Nucl.\
Phys.\ B {\bf 667}, 90 (2003) [arXiv:hep-th/0302106].
}
\lref\AlexandrovUH{ S.~Alexandrov, ``Backgrounds of 2D string
theory from matrix model,'' arXiv:hep-th/0303190.
}
\lref\AlexandrovUT{ S.~Alexandrov, ``Matrix quantum mechanics and
two-dimensional string theory in non-trivial backgrounds,''
arXiv:hep-th/0311273.
}

\lref\SeibergBJ{ N.~Seiberg and S.~H.~Shenker, ``A Note on
background (in)dependence,'' Phys.\ Rev.\ D {\bf 45}, 4581 (1992)
[arXiv:hep-th/9201017].
}
\lref\ShenkerUF{ S.~H.~Shenker, ``The Strength Of Nonperturbative
Effects In String Theory,'' RU-90-47 {\it Presented at the Cargese
Workshop on Random Surfaces, Quantum Gravity and Strings, Cargese,
France, May 28 - Jun 1, 1990} }
\lref\KarczmarekPV{ J.~L.~Karczmarek and A.~Strominger, ``Matrix
cosmology,'' arXiv:hep-th/0309138.
} \lref\McGreevyEP{ J.~McGreevy, J.~Teschner and H.~Verlinde,
``Classical and quantum D-branes in 2D string theory,'' JHEP {\bf
0401}, 039 (2004) [arXiv:hep-th/0305194].
}
\lref\giddings{ S.~B.~Giddings and A.~Strominger, ``Exact black
five-branes in critical superstring theory,'' Phys.\ Rev.\ Lett.\
{\bf 67}, 2930 (1991).
}

\lref\DabholkarWN{ A.~Dabholkar and C.~Vafa, ``tt* geometry and
closed string tachyon potential,'' JHEP {\bf 0202}, 008 (2002)
[arXiv:hep-th/0111155].
}
\lref\McGreevyKB{ J.~McGreevy and H.~Verlinde, ``Strings from
tachyons: The c = 1 matrix reloaded,'' JHEP {\bf 0312}, 054 (2003)
[arXiv:hep-th/0304224].
}
\lref\cta{ Y.~Okawa and B.~Zwiebach, ``Twisted Tachyon
Condensation in Closed String Field Theory,''
arXiv:hep-th/0403051.
}
\lref\ctb{ M.~Headrick,
arXiv:hep-th/0312213.
} \lref\ctc{ S.~Sarkar and B.~Sathiapalan, ``Closed string
tachyons on C/Z(N),'' arXiv:hep-th/0309029.
}
\lref\bta{ T.~Suyama, ``On decay of bulk tachyons,''
arXiv:hep-th/0308030.
}

\lref\rabinovici{
S.~Elitzur, A.~Forge and E.~Rabinovici,
``Some global aspects of string compactifications,''
Nucl.\ Phys.\ B {\bf 359}, 581 (1991).
}

\lref\btb{ A.~A.~Tseytlin, ``Magnetic backgrounds and tachyonic
instabilities in closed string  theory,'' arXiv:hep-th/0108140.
} \lref\jlkas{ J.~L.~Karczmarek and A.~Strominger, ``Closed string
tachyon condensation at c = 1,'' arXiv:hep-th/0403169.
}


\lref\RST{J.G.~Russo, L.~Susskind, and L.~Thorlacius, Phys.~Lett.
B292 (92) 13.} \lref\CGHS{C.G. Callan, S.B. Giddings, J.A. Harvey,
and A. Strominger,  Phys. Rev. D45 (92) R1005.}
\def\apm{\alpha^\prime }
\overfullrule=0pt
\def\Title#1#2{\rightline{#1}\ifx\answ\bigans\nopagenumbers\pageno0\vskip1in
\else\pageno1\vskip.8in\fi \centerline{\titlefont #2}\vskip .5in}
 
scaled\magstep3 
 
scaled\magstep3 
 
scaled\magstep3


\newcount\figno
\figno=0
\def\fig#1#2#3{
\par\begingroup\parindent=0pt\leftskip=1cm\rightskip=1cm\parindent=0pt
\baselineskip=11pt \global\advance\figno by 1 \midinsert
\epsfxsize=#3 \centerline{\epsfbox{#2}} \vskip 12pt {\bf Fig.\
\the\figno: } #1\par
\endinsert\endgroup\par
}
\def\figlabel#1{\xdef#1{\the\figno}}
\def\encadremath#1{\vbox{\hrule\hbox{\vrule\kern8pt\vbox{\kern8pt
\hbox{$\displaystyle #1$}\kern8pt} \kern8pt\vrule}\hrule}}

%
\Title{\vbox{\baselineskip12pt \hbox{hep-th/0411174}}} {\vbox{
\centerline {Black Hole Non-Formation in the } \centerline{ Matrix Model}}} \centerline{Joanna L.~ Karczmarek, Juan
Maldacena and
Andrew Strominger} 

\vskip.1in \centerline{\bf Abstract} {The leading classical
low-energy effective actions for two-dimensional string theories
have solutions describing the gravitational collapse of shells of
matter into a black hole. It is shown that string loop corrections
can be made arbitrarily small up to the horizon, but $\apm$
corrections cannot. The matrix model is used to show that typical
collapsing shells do not form black holes in the full
string theory. Rather, they backscatter out to infinity just before
the horizon forms. The matrix model is also used to show that the
naively expected particle production induced by the collapsing
shell vanishes to leading order. This agrees with the string
theory computation. From the point of view of the effective low
energy field theory this result is surprising and involves a
delicate cancellation between various terms.
 }

 \Date{}

\listtoc\writetoc
\newsec{Introduction}
Despite much recent progress, a number of interesting questions
concerning two-dimensional string theories and their matrix model
solutions remain unanswered.  
Among them is the seemingly simple question
of whether or not the
theories contain black holes. If they do, we would like to understand the
quantum behavior of the black holes. If they do not,
we would like to understand why not.

The continuum 2d string theory has an exact classical solution
given by an $SL(2,R)_k/U(1)$ worldsheet CFT \WittenYR. For large
$k$ this CFT has an unambiguous interpretation as a spacetime
black hole \refs{\rabinovici,\MandalTZ,\WittenYR}\foot{More
precisely for large $k$ it is a factor of the CFT for the near
extremal black fivebrane in critical string theory \giddings. For
small $k$, which is relevant to the present work, the black hole
interpretation is less clear, as will be discussed in the
concluding section.}. Moreover, for the small values of $k$
relevant to 2d string theory, an exact matrix model (significantly
involving the non-singlet sector) which generalizes this classical
solution to the full quantum theory is known \kkk. On the face of
it these facts suggest that black holes indeed arise in 2d string
theories much as in their higher dimensional cousins. Furthermore,
the computations of \kkk\ show not only that the $SL(2)/U(1)$
black hole arises in the non-singlet sector of the matrix model
but also that there is a wider family of black holes than we would
have suspected from the low energy effective action, since one can
vary the temperature.

On the other hand, so far there has been no clear indications of
black hole formation in any matrix model scattering amplitudes \ref\MoorePlesser{
G.~W.~Moore and R.~Plesser,
``Classical scattering in (1+1)-dimensional string theory,''                   
Phys.\ Rev.\ D {\bf 46}, 1730 (1992)
[arXiv:hep-th/9203060].
},
\Martinec\
(see \PolchinskiMB\ for an early discussion). There is also no sign
of a large degeneracy of states which would be required for a
microscopic accounting of black hole entropy, or the phase
transition signaling black hole formation.
Finally, the quintessentially thermal nature of black holes
seems at odds with the integrability of the matrix model \sshe .

Since all of these facts involve purely the singlet sector of the
matrix model, all that is known so far is consistent with the
black holes arising {\it only} in the non-singlet sector. Since
the non-singlet sector is much harder to analyze in the Lorentzian
context than the singlet sector, we would like to make sure we are
not missing some black hole configurations that arise in the
singlet sector.

In this paper we address this problem by trying to make black
holes by collapsing the tachyonic or axionic matter. It is shown
that the low-energy effective action indeed has solutions in which
such collapse occurs in a region where string loop corrections can
be made arbitrarily small. 
This suggests that the problem of black hole formation 
is a question for classical string theory or
equivalently worldsheet conformal field theory.
On the other hand, $\apm$ corrections
can never be controlled, so a full classical string analysis is
needed. Using the matrix model, we show that a typical collapsing
pulse does {\it not} form a black hole. Rather it bounces off of
the would-be event horizon just before the black hole is about to
form, and the matter is backscattered back out to infinity!

We further analyze the problem of particle creation  by the
collapsing shell. Naively, one expects a rising tail of outgoing
Hawking radiation even before a black hole forms.  A  matrix model
computation is used to show that the leading term (at early
times) of tachyon particle production on ${\cal I}^+$
vanishes. A consistent picture emerges from a careful analysis
of the spacetime effective action, which reveals several sources
of particle production in addition to the Hawking radiation.
For appropriate numerical values of the effective action coefficients,
these different contributions can cancel, in agreement with
the matrix model computation.
We also perform the exact (in $\alpha'$) string theory scattering
amplitude computation and show that the amplitude vanishes here
(this is the bulk $(2,2)$ amplitude studied in \refs{\dgik,\Francesco}.

Our results are certainly consistent with the idea that there are no
black holes in the singlet sector of the
matrix model. At the same time, our investigations have
uncovered more curious behavior indicating there is much we do not
understand about the matrix model. As such we feel it is too early to jump
to conclusions. Further discussion can be found at the end of this paper.

This paper is organized as follows: In section 2 we describe the
collapse of a matter shell in the 2 dimensional effective theory
for 0B string and discover that while other corrections are under
control, the curvature on the would-be horizon is of order one and
thus the effective description breaks down.  In section 3 we
describe the same pulse in matrix theory and discover that most of
the energy does in fact escape to infinity at parametrically the
same retarded time at which the horizon would have formed. In
section 4 we switch gears and begin to study early time particle
creation in the collapsing pulse.  We consider various sources of
early time outgoing stress energy for both the tachyon and the
axion in effective theory.   In 4.4 we study the related string
theory S-matrix element. In section 5, we study this outgoing
stress energy in the matrix model at $\mu=0$ using the free
fermion formulation. Finally, in section 6, particle creation due
to a very energetic incoming state is considered, with
implications for possible thermality of the outgoing state. We
close with discussion in section 7.

Related independent work has recently appeared in \FriessTQ.
Previous work on particle creation in the matrix model includes
\refs{\aj \finn \MukhopadhyayFF -\aaj}.

\newsec{Spacetime gravitational collapse }

The leading bosonic terms of the low-energy effective action for
type 0B string theory are \eqn\seff{S_{eff}={ 1 \over 2 \pi}\int
d^2x \sqrt{-g}\bigl(e^{-2\Phi}\bigl(16+R +4(\nabla \Phi)^2-\half
(\nabla T)^2 +2T^2)-\half(\nabla C)^2+.....\bigr).} In this
expression $\alpha^\prime=\half$, $T$ is the NS-NS sector
"tachyon" and $C$ is the RR axion. The position-dependent string
coupling is given by the dilaton \eqn\mmn{g_s=e^\Phi.}
The corrections to \seff\ involve higher derivatives, more powers
 of $g_s$
and/or higher nonlinear dependence on $T$. To all orders in
perturbation theory there is a RR shift symmetry which prohibits
the axion $C$ from appearing without a derivative. The classical
terms in the action all scale like $\lambda^2$ under
$\Phi\to\Phi-\ln \lambda,~~~~C \to \lambda C$, so that the
classical limit is \eqn\csc{C \sim e^{-\Phi}\sim {1 \over g_s}\to
\infty.} In this section we study black hole formation using the
effective action \seff, and also discuss when and how the approximation
\seff\
to the exact theory becomes unreliable.

\subsec{The tachyon}

  The question of whether or not gravitational tachyon collapse occurs
  at $c=1$ was discussed
  in the early days of the bosonic matrix
  model, which does not have the axion. A naive argument that there is no reliable approximation
  in which the tachyon field can be seen to form an event horizon goes
  as follows.  In order for an event horizon to form, the metric and or dilaton
  must be of order one. The equations of motion following from
  \seff\ allow this to happen only if the tachyon $T$ is also of
  order one.  However if the tachyon is of order one, its self
  interactions are important, and \seff\ can not be trusted. Hence
  one cannot be sure that gravitational collapse of tachyons can
  really occur.

This argument is a bit too fast because we should also consider the
possibility of making $T$ small by spreading out the pulse. However
the problem of tachyon collapse is still difficult to analyze in
part because of the complicated tachyon-dilaton interactions
present in \seff, as well as the fact that the stress tensor is not positive
definite even at leading order. We therefore turn to the problem of axion
collapse which turns out to be much simpler.

\subsec{The axion}

Setting the tachyon to zero and scaling the axion as in \csc, the
action \seff\ becomes precisely the classical CGHS \CGHS\ action
coupling dilaton gravity to conformal matter.\foot{This is not
exactly a coincidence, since the CGHS theory was derived from the
linear dilaton theory in the NS fivebrane throat.}  It is
convenient to choose conformal gauge $g_{\mu \nu} =
e^{2\rho}\eta_{\mu \nu}$ or, in light-cone coordinates,
\eqn\ththree{ ds^2=- e^{2 \rho}dx^+dx^-.}
We then have $R=8 e^{-2 \rho} \partial_+ \partial_- \rho$ and the
equations of motion become
\eqn\thfour{\eqalign{ \Phi &: \qquad e^{-2(\Phi+\rho)} \left[ -4
\p_+ \p_- \Phi +
 4 \p_+ \Phi \p_- \Phi +
        2\p_+ \p_- \rho + 4e^{2 \rho} \right] = 0, \cr
\rho &:\qquad   e^{-2 \Phi} \left[ 2 \p_+ \p_- \Phi - 4 \p_+ \Phi
\p_- \Phi
                     - 4 e^{2 \rho} \right] =0,\cr
                     C&:\qquad  \p_+\p_-C=0. \cr }}
 Since we have gauge fixed
$g_{++}$ and $g_{--}$ to zero we must also impose their equations
of motion as constraints. These are
\eqn\thsix{\eqalign{ e^{-2 \Phi} ( 4 \p_\pm \rho \p_\pm \Phi - 2
{\p_\pm}^2 \Phi ) &= -T^C_{\pm \pm }, \cr T^C_{\pm\pm}\equiv \half
\p_\pm C\p_\pm C&~~. \cr }}

We are interested in the general infalling  axion solution which
is \eqn\aeom{C(x^+,x^-)=C(x^+).}
 Further simplifications are obtained using the residual gauge
freedom to choose Kruskal gauge
\eqn\kgc{\Phi=\rho,} so that
\eqn\dsa{ds^2=-e^{2\Phi}dx^+dx^-.} In this gauge the linear
dilaton vacuum is $e^{-2\Phi}=-4x^+x^-$ with the Kruskal
coordinates running over the range $-\infty<x^-<0$, $0<x^+<
\infty$. The ++ constraint equation is
\eqn\zio{\p_+^2e^{-2\Phi}=-T^C_{++}.} The general solution of
\thfour\ and \zio\ is then
\eqn\iiy{e^{-2\Phi}=-4x^+x^--\int^{x^+}dy^+\int^{y^+}dz^+T^C_{++}(z^+).}

\fig
{Penrose Diagram for the formation of a black hole in 2d string theory.}
{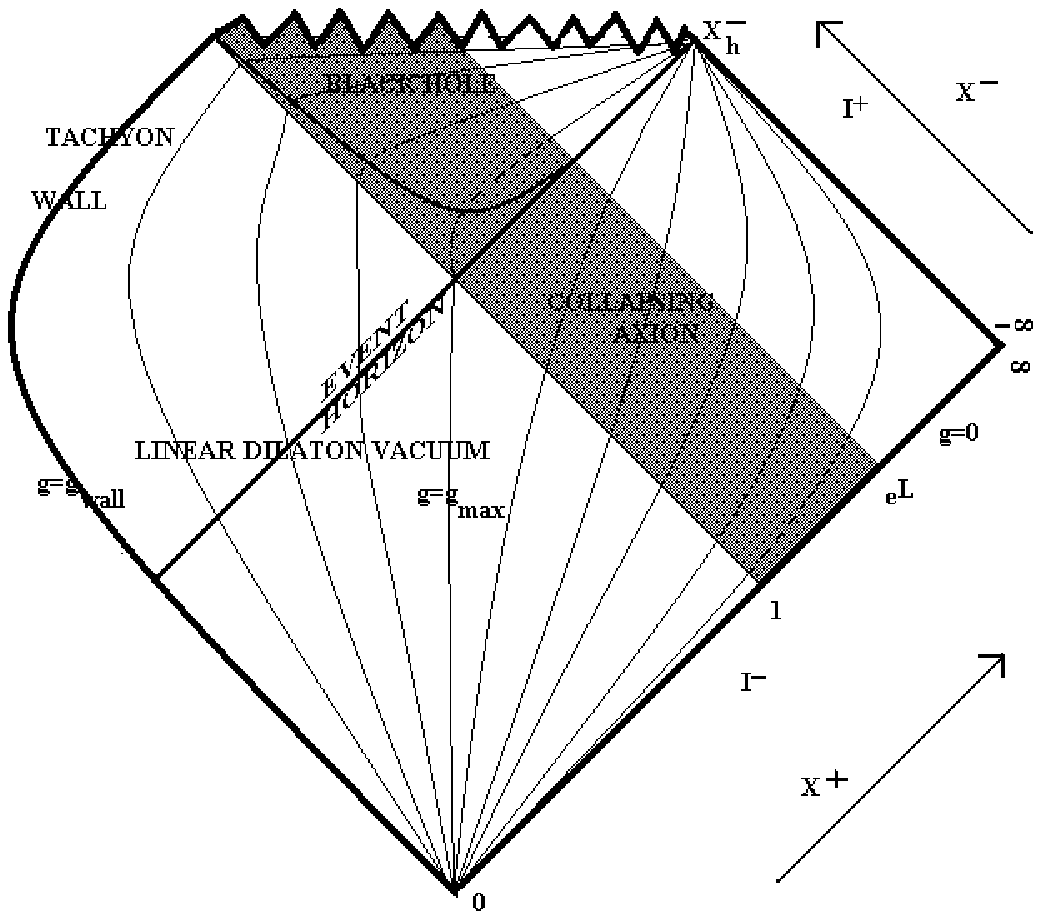}{0.0truein}

Let us send in a pulse so that $T^C_{++}$ is non-vanishing only
between $x^+=1$ and $x^+=e^L$. (This is an interval of length $L$
in the asymptotically inertial coordinate $\sigma^+=\ln x^+$.) The
total mass of the pulse as computed on ${\cal I}^-$ is \eqn\mpl{M=
\int dx^+ x^+ T^C_{++}~ .} The general solution \iiy\ for such a pulse
is depicted in figure 1. An event horizon, defined as the
boundary of the region from which light rays can escape to ${\cal
I}^+$, forms at some value of $x^-$ which we denote $x^-_H$. There
is also an apparent horizon, defined as the line along which the
dilaton gradient turns from spacelike to null so that
$\p_+e^{-2\Phi}=0$.\foot{In the derivation of \seff\ from
dimensional reduction, $e^{-\Phi}$ originates as the radius of the
higher dimensional spheres. Hence from this perspective the
apparent horizon is the boundary of the region of trapped spheres
\RST.} Classically the apparent horizon is always inside the event
horizon. It crosses the point $(e^L, x^-_H)$, i.e. the
intersection of the event horizon with the top of the pulse. It
follows that \eqn\rfd{\eqalign{ 0&=\p_+e^{-2\Phi}(e^L,x^-_H)\cr
&=\int_1^{e^L}dx^+\p_+^2
e^{-2\Phi}(x^+,x^-_H)+\p_+e^{-2\Phi}(1,x^-_H)\cr &
=-\int_1^{e^L}dx^+ T^C_{++}(x^+) +\p_+e^{-2\Phi}(1,x^-_H).}}
Equivalently, using the fact that below the pulse we have the
linear dilaton vacuum, \eqn\dsr{-{
4x^-_H}=\int_1^{e^L}dx^+T^C_{++}.} We note that the maximum value of
the string coupling relevant to this computation
\eqn\dza{g_H^2=-{1 \over 4 x_H^-}} occurs where the bottom of
the pulse intersects the event horizon.

\subsec{Corrections}

The first issue is whether or not semiclassical perturbation
theory breaks down. In order that it not break down we need the
string coupling $e^\Phi$ to be small everywhere outside the event
horizon and in the non-vacuum region above the bottom of the pulse
($x^+>1$). Causality\foot{Of course it is not completely clear to what
extent spacetime causality follows from the matrix model.}
forbids a strong coupling region inside the
horizon from affecting horizon formation. The maximum value of
the string coupling in this region is given by \dza, so in order
for perturbation theory to be good we need the integral in \dsr\
to be large. This is obviously easy to arrange just by making
$T^C_{++}$ large. That is, if we throw in enough energy the black
hole can be formed in a region of weak string coupling where
quantum corrections to \seff\ can be made arbitrarily small.

However we still need to worry about higher-derivative
$\alpha^\prime$ corrections to \seff.\foot{For the tachyon self
interaction terms such as $e^{-2\Phi}T^3$ must also be considered.
The absence of these for the axion facilitates the analysis.} One
example of such a classical term (consistent with \csc ) is
\eqn\vfs{\int d^2x\sqrt{-g}e^{2\Phi}(\nabla C)^4.} However this
term identically vanishes for the solution \aeom\ because we have
taken $\p_-C=0$. To get something non-zero we need a term
like
\eqn\dsko{\int
d^2x\sqrt{-g}e^{2\Phi}(\nabla C \cdot \nabla \Phi)^4~.} This would
give a correction term to the stress tensor of the form
\eqn\wsz{\Delta T^C_{++} \sim e^{2\Phi} (\p_+C)^4.}
 We need this to be small relative to the leading term \dsr. This is also
easy to arrange. Choose constant $\p_+C=f$ inside the pulse. Then we have from
\dsr\ and \dza\ \eqn\jua{ f={\sqrt{2} \over g_H\sqrt{e^L-1}}.} The
correction term obeys \eqn\dzsx{
{\Delta T^C_{++} \over T^C_{++} }
<2 g_H^2(e^L-1)f^2={4 \over (e^L-1)}~.} 
So we need only take $e^L-1 >>1$ in order
for this to be a small correction. This shows that corrections of
the type \wsz\ cannot prevent black hole formation in all cases.

However there is another type of $\apm$ correction which can not
be controlled: couplings to curvature. One such term is
\eqn\dsz{R(\nabla C)^2.} Since $R$ is of order one in string units
at the horizon, such terms may have a large effect at the horizon.
Hence we cannot be sure whether or not a black hole really forms
in the full string theory.

We may also consider how the background tachyon $T$ affects the
story. In the preceding discussion we have set it to zero. This is
consistent with the leading equations of motion, so if there is a
tachyon tadpole at some order it would be suppressed. However we
could also consider the effects of the tachyon wall, $i.e.$ a
tachyon vev of the form \eqn\tru{T \sim{\mu \over
\sqrt{-x^+x^-}}.} This produces a wall at the region where $T$ is
of order one. One expects the axion to be reflected from the wall.
This could be described for example by an interaction term
\eqn\ssi{\int d^2x\sqrt{-g} e^{ T} (\nabla C)^2 ~,} which effectively imposes
a reflecting boundary condition for
$C$ in the region where the tachyon is rapidly varying.
The location of the tachyon wall can be characterized by
$g_{wall}$, the value of the string coupling when $T \sim 1$. By
taking \eqn\raw{g_{wall}\gg g_H,} we can arrange for the tachyon
wall to be far behind the region where the event horizon forms.
Hence all effects of the tachyon vev on events leading up to the
formation of the event horizon are suppressed by powers of ${g_H
\over g_{wall}}$, which again can be made arbitrarily small. Of
course if we take the tachyon wall to be outside the region where
the event horizon would form, the collapsing pulse should reflect
back to infinity rather than make a black hole.

In conclusion, due to uncontrollable $\apm$ corrections involving
the curvature near the horizon, the question of whether or not
black hole formation occurs cannot be answered within low energy
field theory. However, since string loop corrections can be
suppressed, it is a question for classical string theory or
equivalently worldsheet conformal field theory. A construction of a
worldsheet CFT describing black hole formation would certainly be of great
interest.

\newsec{Matrix model picture}

The classical limit \csc\ in spacetime corresponds to the Fermi
liquid approximation \PolchinskiUQ\ in the free fermion
formulation of 2D string theory. In this section we describe the
collapsing pulses of the previous section as finite perturbations
of the Fermi surface and analyze their behavior. We will be
working in the matrix model of the type 0B string.

\subsec{Collapsing pulse as a Fermi sea perturbation }
The first step is to express the pulse in the standard linear
dilaton Liouville theory coordinates \eqn\ttg{t^\pm = t \pm \ln
\lambda,} where $\lambda$ is the matrix model eigenvalue governed
by the Hamiltonian\foot{For type 0 (bosonic) theory this is
$\alpha^\prime=\half$ ($\alpha^\prime=1$).}
\eqn\mmh{H=\half (\p_t\lambda)^2-\half
\lambda^2.} The asymptotic form of the
dilaton is \eqn\wsa{\Phi = -2\ln\lambda =t^- - t^+} and $ds^2 = -
dt^+ dt^-$. Their relation to the Kruskal coordinates $x^\pm$ of the
previous section is \eqn\krf{x^\pm =\pm\lambda^2 e^{\pm
2t}=\pm\half e^{\pm 2t^\pm}.}

Thus, a constant ${\p \over \p x^+ }C =  f$ becomes $\p_{+}C = f
e^{2t^+}$ over the region
\eqn\pulsriv{ \p_+ C = fe^{2t^+} ~~~\half \ln 2 <t^+< \half \ln 2
+ {L \over 2}~ .}
Notice that the pulse height reaches a  maximum value
\eqn\iom{|\p_+ C| \sim 2f e^{L}}
which is exponentially large in $L$. It can be seen that this is
generically the case. Actually in the matrix
model it is awkward to describe pulses with $\int\p C \neq 0$.
It is not hard to arrange for this integral to  vanish. In these
cases one finds that
the pulse reaches values of order $e^{L}$
of either sign.

According to \refs{\TakayanagiSM,\DouglasUP} a RR pulse corresponds to a
left-right antisymmetric fluctuation of the two branches of the
Fermi sea which we denote $\eta_{RR}$.
The correspondence involves the leg-pole
transform \eqn\legpole {\partial \eta_{RR}(t^+) = \int dy K(y)
\partial_{+}C(t^+ - y)} where the kernel is given by
\eqn\kernel{
K(y) = \int {dk \over 2\pi} \Big ({\pi \over 2} \Big )^{-ik/8}
{\Gamma(1/2 + ik/2)  \over \Gamma(1/2 - ik/2) } e^{iky} = {z  }
J_0(z)~,} with \eqn\zdf{ z \equiv 2 \Big ({\pi \over 2}
\Big)^{1/8} e^{-y }~. } This changes the shape of the pulse
mostly on scales of order $\apm$. In particular, as can be shown
quite generally,
 the Fermi sea fluctuations corresponding to the spacetime pulses
under discussion themselves have magnitude $e^{L}$ and are of either sign.
 As an illustration, a graph of the  pulse
\eqn\pulseDeriv{\eqalign{ \p_+ C &= fe^{2t^+} ~~~\half\ln 2<t^+<
\half \ln \big({e^L + 1 }\big)
\cr \p_+ C &= -f e^{2t^+}~~~\half \ln \big(
 e^L + 1 \big)<t^+<
\half \ln 2 + { L \over 2} }}
(which has $\int\p C = 0$), together with
its leg-pole transform is given in figure 2.
\fig
{Derivative $\p_+C$ of the axion pulse (filled) and its leg pole
transform as a function of the inertial coordinate $t^+$ for $L=5$.}
{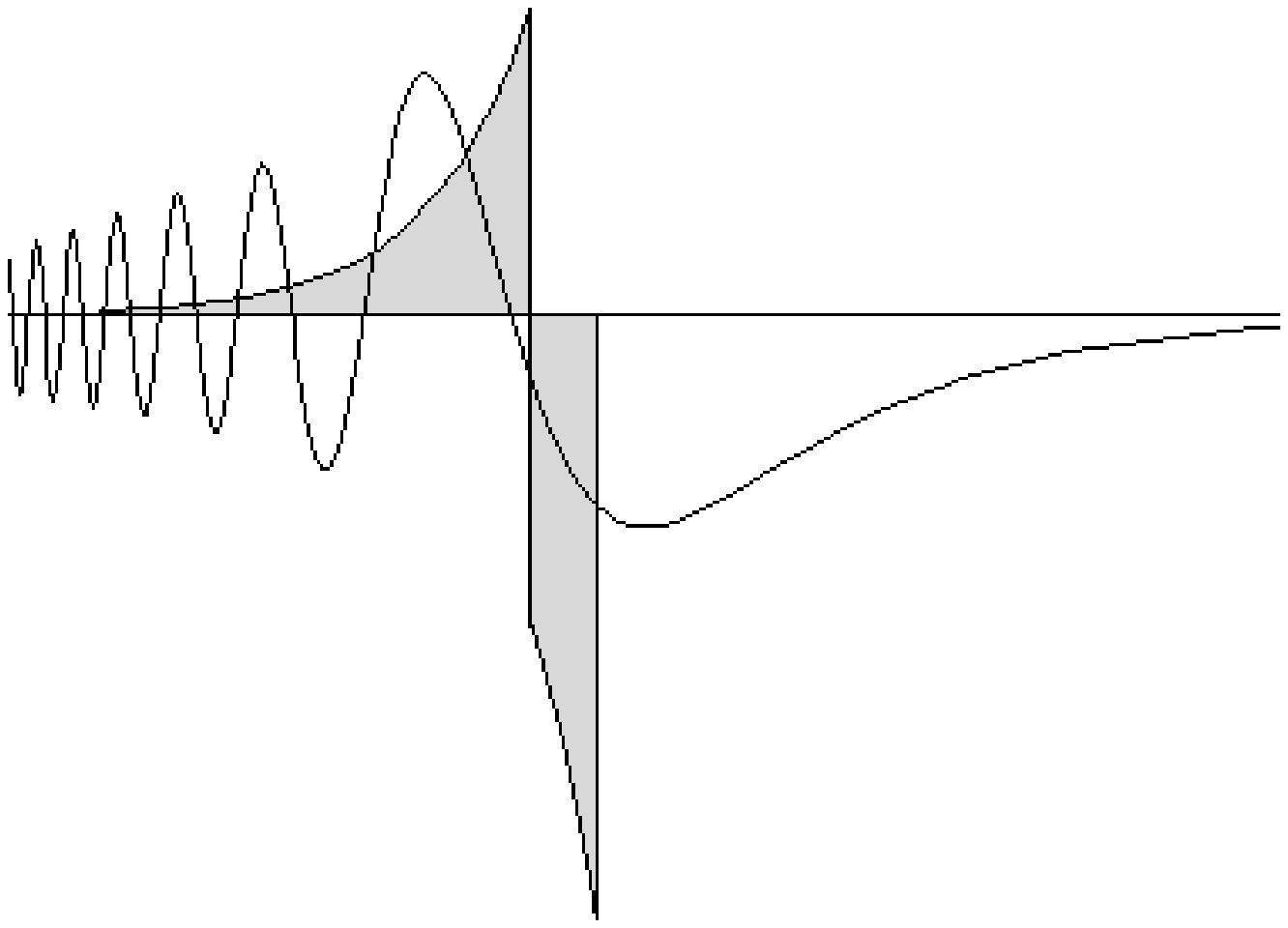}{0.0truein}

The incoming fluctuations live on the lower branch of the
right hyperbola and on the upper branch of the left hyperbola.
These are given by \GukovYP
\eqn\ed{
p^r_+(\lambda,t) = -\sqrt{\lambda^2 - g_{wall}^{-1}} - {1 \over 2 \lambda}
\partial_{t^+}\eta_{RR}~,
}
and
\eqn\eded{
p^l_+(\lambda,t) = \sqrt{\lambda^2 - g_{wall}^{-1}} + {1 \over 2 \lambda}
\partial_{t^+}\eta_{RR}~,}
which can be expanded as
\eqn\ftop{p^{r/l}_+(\lambda,t)
=-\lambda+{1 \over 2 g_{wall} \lambda} \mp {\partial_{t^+}\eta_{RR}
\over 2 \lambda}.}
The pulse spills from right to left (and, correspondingly, from left
to right), if at some point, $\partial_{t^+}\eta_{RR} > g_{wall}^{-1}$
($\partial_{t^+}\eta_{RR} < -g_{wall}^{-1}$).  From \iom,
both of these conditions are satisfied somewhere as long as
$f e^L > g_{wall}^{-1}$.  With \jua, this implies
$g_H < e^{L/2} g_{wall}$, consistent with \raw, the condition
that the tachyon wall is well behind the region where the event
horizon forms in effective theory.
Note that making the pulse wider
(making $L$ large) keeping $g_H$ fixed only makes this
inequality easier to satisfy.

\subsec{ Outgoing radiation on ${\cal I}^+$}

In this subsection we will see that, in the matrix model,
the spilled-over peaks of the pulse (as well as the minima)
are reflected off the strongly coupled region early and
begin to arrive at ${\cal I}^+$ at a retarded
time which coincides parametrically with the time for semiclassical
black hole formation.
We shall further see that the majority of the pulse returns to  ${\cal  I}^+$
in a relatively short time.  We will assume that
$g_H \ll e^{L/2} g_{wall}$, or that the fluctuations of the incoming
pulse are large compared to $g_{wall}^{-1}$.

Let the generic bottom tip of the pulse be at
\eqn\dzao{
\partial_{t^+}\eta_{RR} = \epsilon \gg g_{wall}^{-1}~}
This point moves along the hyperbola
  \eqn\dook{\lambda^2-p^2=\epsilon,}
  according to
  \eqn\erfu{\lambda=\sqrt{\epsilon}\cosh (t + t_0).}
Using \ttg\ and \krf, we find that in the past, the point is at
\eqn\www{
x^+ = {\epsilon \over 8} e^{-2t_0}
}
In the future, we have
\eqn\vvv{
x^- = {\epsilon \over 8} e^{2t_0} \sim {\epsilon^2 \over x^+}
}
In the past, the Kruskal coordinate of the infalling tip of the pulse
is is $x^+ \sim e^L$, thus, using \jua\ and \iom, we obtain that
\eqn\uuu{
x^- \sim g_H^{-2} \sim x_H^-~.
}
A parallel computation can be performed for the spilling over
peaks of the pulse, with the same conclusion.

Ordinarily one expects that retarded times for pulse reflection diverge for
$g_{wall}^{-1} \to 0$ as the tachyon wall is taken to infinity, while \uuu\ is
independent of $g_{wall}$.  What is happening is that the tips of the pulse
are reflected at a time determined by their energy alone. For energetic pulses
obeying the condition \dzao\ the value of $g_{wall}$ is not relevant.
Moreover, the tip of the pulse emerges on ${\cal I}^+$
where the black hole horizon is in semiclassical theory.

According to \vvv, portions of the incoming pulse on trajectories
with $e.g.$
energy $\half \epsilon_0$ (either positive or negative)
 come out a time which is later  by
$\ln 2$ (in inertial coordinates).
Hence the bulk of the pulse will return to ${\cal I}^+$ over
 a time period which is of order one. Ignoring for a
 moment the effects of the leg pole
 transform (see below), the  observer on ${\cal I}^+$ will simply
see a pulse that
is reflected off of the surface where the black hole would have formed in
the semiclassical theory. Hence the $\apm$ corrections act like a shield
preventing the incoming pulse from ever getting inside its Schwarzchild
radius.

The semiclassical picture is distinctly different. Hawking
radiation, consisting mainly of quanta with energies of order one
in string units, begins to appear on ${\cal I}^+$ at retarded
times of order $x^-_H\sim 1/g_H$.  This radiation continues until
the black hole is fully evaporated, which takes a time of order
the total energy $E$ of the incoming pulse, assuming a temperature and
greybody factor of order one.  Hence the energy
return takes a parametrically longer time in this picture than it
does in the exact matrix model. (See \Martinec\ for a related
discussion.)

Now we show that the matrix model  picture is not substantially affected by
the leg pole transform. The transform \legpole\ relates
an outgoing collective field quantum
of energy $\omega$ to that of the axion by the phase shift
\eqn\rabi{e^{i\omega t}\to {\Gamma(1/2 + i\omega/2 )  \over \Gamma(1/2 -
    i\omega/2) }e^{i\omega t}.}
For $\omega$ order one in string units this implies a time delay of order
the string time. For large $\omega$ we have from Stirling's formula
\eqn\rbi{e^{i\omega t}\to e^{i\omega (t+\ln \omega)}.}
Hence the leg pole factor leads to a time delay of order $\ln \omega$ for
highly energetic quanta. Representing the pulse \pulsriv\ by a coherent
quantum state, one finds that a typical quantum has an energy
which goes at most
a power of the total energy $E$. Hence the time delays will go at most like
$\ln E$, and the energy is returned over a time interval which is short
compared to the interval $E$ expected from the black hole picture.

\newsec{Particle Production in the Spacetime Picture}

An infalling pulse of matter perturbs the quantum state of the
outgoing matter, and leads to the quantum pair production of
outgoing matter. There are several sources of this pair
production. In this section
we will compute the pair production in the spacetime picture, and
in the next we will compare it to a matrix model computation.
In addition to Hawking radiation, which exists for both the tachyon
and the axion, we will also compute contributions due to a time dependent
mass (for the tachyon) and a higher order interaction (for the axion and
the tachyon).  Additional outgoing particle flux  arises from 
the $T(\p C)^2$ interaction,
but we will not consider those in this section,
as the resulting flux has a somewhat different form from the
gravitational one.

\subsec{Particle creation by an infalling pulse}
  The most familiar form of  pair production is the
Hawking radiation due to the mismatch of time coordinates on
${\cal I}^+$ and ${\cal I}^-$.  This occurs for an infalling pulse
even if a black hole is not formed, but is exponentially small at
early retarded times. We wish to compute the leading exponential
correction to the ${\cal I}^+$ stress tensor for a massless
particle, which may be either the axion or the tachyon, governed
by the conformal gauge action \eqn\doj{{1 \over 2 \pi}\int d^2u \p_+S\p_-S.}

The metric conformal factor vanishes in the far past in the
coordinates \eqn\crd{t^\pm=\pm \half \ln (\pm 2 x^\pm)} and is
given by
\eqn\fdx{\rho = \Phi+t^+-t^-.} At a generic point,
transforming from \iiy\ one has
\eqn\esx{2\rho=-\ln(1-U(t^+)e^{2t^-})\sim U(t^+)e^{2t^-},} where
\eqn\ews{U(t^+)=e^{-2t^+} \int_{-\infty}^{t^+}ds^+e^{2s^+}
\int_{-\infty}^{s^+}du^+e^{-2u^+}T_{++}(u^+).} For a finite
duration pulse $U$ obeys
\eqn\was{~~~~U(\infty)={1 \over 2} \int^{\infty}_{-\infty}dt^+e^{-2t^+}
T_{++}(t^+).} The asymptotic stress tensor of the Hawking
particles on ${\cal I}^+$, as measured in inertial coordinates, is
given by the Schwartzian of the transformation required to set
$\rho=0$ on ${\cal I}^+$. This is easily seen to be, again to
leading order, \eqn\poh{T_{HR--} = { 1 \over 6} U(\infty)e^{2t^-}.}

\subsec{A time dependent mass} In this
subsection we consider the general problem of particle production
by a time-dependent mass term, \eqn\doj{{1 \over 2 \pi}\int
d^2u(\p_+S\p_-S-m^2(u^+,u^-)S^2),} where we take $m^2$ to vanish
on ${\cal I}^\pm$. Such a term is forbidden for the axion but is
actually present, with a form given below, for the tachyon.  In
general this time dependent mass will lead to particle production.
We will compute the corresponding two point function in the in
vacuum \eqn\fgk{ \Delta (s;t)\equiv \langle in | S(s^-)S(t^-)| in
\rangle>} to first order in $m^2$ on ${\cal I}^+$. Expanding in
powers of $m^2$ \eqn\dxq{\Delta=\Delta_0+\Delta_1+\cdots} we find
\eqn\rih{{\p^2 \over\p  s^+\p
s^-}\Delta_1(s;t)=-m^2(s)\Delta_0(s;t),} where \eqn\rwa{
\Delta_0(s;t)=\ln (s^--t^-).} There is a similar equation obtained
by the interchange of $s$ and $t$. The solution of \rih\ is
\eqn\yyo{\Delta_1(s;t)= -\int_{-\infty}^{s^+}du^+
\int_{-\infty}^{s^-}du^-m^2(u^+,u^-)\Delta_0(u;t)+s\leftrightarrow
t,} Defining  the null integral of $m^2$
\eqn\yio{m^2(u^-)=\int_{-\infty}^\infty du^+m^2(u^+,u^-),} \yyo\
can be written on ${\cal I}^+$ ($s^+\to\infty$, $t^+\to \infty$)
as \eqn\rtou{\Delta_1(\infty,s^-;\infty, t^-)
=-\int_{-\infty}^{s^-}du^-{m^2(u^-) \ln (u^--t^-)}-
\int_{-\infty}^{t^-}du^-{m^2(u^-) \ln (u^--s^-)}.} We are
particularly interested in the first order correction to the
${\cal I}^+$ stress tensor, which is the $s\to t$ limit of the
second derivative of $\Delta_1$ \eqn\psa{T_{1--}(t^-)=  {\p^2
\over 2 \p s^- \p t^-}\Delta_1(\infty,s^-;\infty, t^-)\big|_{s^- =
t^-} =\half \p_-m^2(t^-).}

Now we compute $m^2$ for the tachyon.  In the linear dilaton
vacuum, the wave equation for the rescaled tachyon
\eqn\fai{S=e^\Phi T} is that of a massless field. However in a
background with a nontrivial metric and dilaton the tachyon
acquires a mass. To leading nontrivial order in $\apm$ the
effective mass is \eqn\sact{ m^2(u^+, u^-)=-4 +(\nabla \Phi)^2
-\nabla^2\Phi.} Fixing conformal gauge and expanding
\eqn\pef{\Phi=t^--t^++\phi,} one finds to linear  order
that\foot{We are using here the conventional form of the effective
action which does not contain for example a
$T^2e^{-2\Phi}\p_-\p_+\Phi$ term. } \eqn\jjh{m^2=
-4(2\rho-\p_-\phi+\p_+\phi -\p_+\p_-\phi).} Imposing the gauge
condition $\rho = \phi$, and using the solution
\esx\ for $\phi$ and $\rho$,  to leading order at early retarded
times one has,\foot{A further correction to the mass might arise
if there as an $e^{-2\Phi}RT^2$ term in the effective action. This
would change the coefficient but not the functional form of the
action.} \eqn\dsx{T_{--}(t^-)=\half \p_-m^2(t^-)=4
U(\infty)e^{2t^-}.} Note that this has exactly the same functional
dependence on the incoming stress tensor as \poh.

In the calculation we have done, the origin of the two types of
particle production seem quite different. One comes from a
Bogolubov transformation on $\cal I^+$, while the other comes from
an explicit time dependent interaction. This difference disappears
in the less familiar null gauge \eqn\metrga{ ds^2 = - dx^+ dx^- +
g_{++} (dx^+)^2 .} In this gauge, unlike conformal gauge,  the
metric approaches unity on ${\cal I}^+$ and there are explicit
interactions between the metric and the massless field $S$.

\subsec{The Quartic Interaction}

In this section we study an
interaction of the form given in \vfs
\eqn\tdi{{1 \over 2\pi}\int d^2u \bigl(\p_+C\p_-C +
Je^{2\Phi-2\rho}\p_+C\p_+C\p_-C\p_-C\bigr)~,}
where J is a constant.  Unlike the mass term in \doj, such a term is not
forbidden for the axion.  There is also an identical term for the
rescaled tachyon field $S$.

We are interested in how the incoming stress tensor $T_{++}$ affects the
outgoing $\p_-c$ two point function. This is governed by  the
equation of motion \eqn\rtio{
\p_+\p_-C+2JT_{++}\p_-\bigl(e^{2t^--2t^+} \p_-C\bigr)=0,} where we
have used \fdx. Repeating the steps of the previous section, one
finds the last term in \rtio\ implies an additional correction to
the two-point function \eqn\rxtu{ \Delta_J (\infty,s^-;\infty,
t^-) =\int_{-\infty}^{s^-}du^-J(u^-)\bigl[{2
\over(u^--t^-)}-{1\over(u^--t^-)^2} +\bigr]~+ ~t^-\leftrightarrow
s^- ,} where \eqn\yyt{J(u^-)=-2Je^{2u^-}\int_{-\infty}^\infty du^+
e^{-2u^+}T_{++}=-4Je^{2u^-}U(\infty).} The stress tensor correction
is then \eqn\psavs{T_{J--}(t^-)=  {\p^2 \over \p s^- \p
t^-}\Delta_1(\infty,s^-;\infty, t^-)\big|_{s^- = t^-} =
-e^{2t^-}U(\infty){16 J\over 3}.} We note that this exactly cancels
the Hawking particle production \poh\ for $J={1 \over 32}$.

\subsec{ Particle production in string theory}

We will now perform the exact string theory computation of 
particle production. Note that the process we are interested in
occurs in the bulk of the two dimensional spacetime, far
from the Liouville wall. Such processes are fairly easy to compute
in string theory, since we can simply use the linear
dilaton conformal field theory. Such amplitudes were computed in
\dgik, \Francesco. In order to make sense of these amplitudes
one needs to consider wavepackets that are very localized,
 as explained in \NatsuumeSP . For example, we can
consider  localized gaussian wavepackets. We are
interested in a process where  an incoming left-moving RR axion
creates a gravitational field which in turn creates two RR
axions. So the net process is one in which we have an incoming
left-moving RR axion and an out going left-moving RR axion and two
right-moving RR ``Hawking'' particles (see figure 3(a)).
\fig{(a) Diagram representing creation of Hawking
radiation. The graviton is not a physical propagating field,
we are just thinking of it as an off shell particle.
(b) Diagram used for computation of the static field created by the
incoming pulse, as studied in \NatsuumeSP} {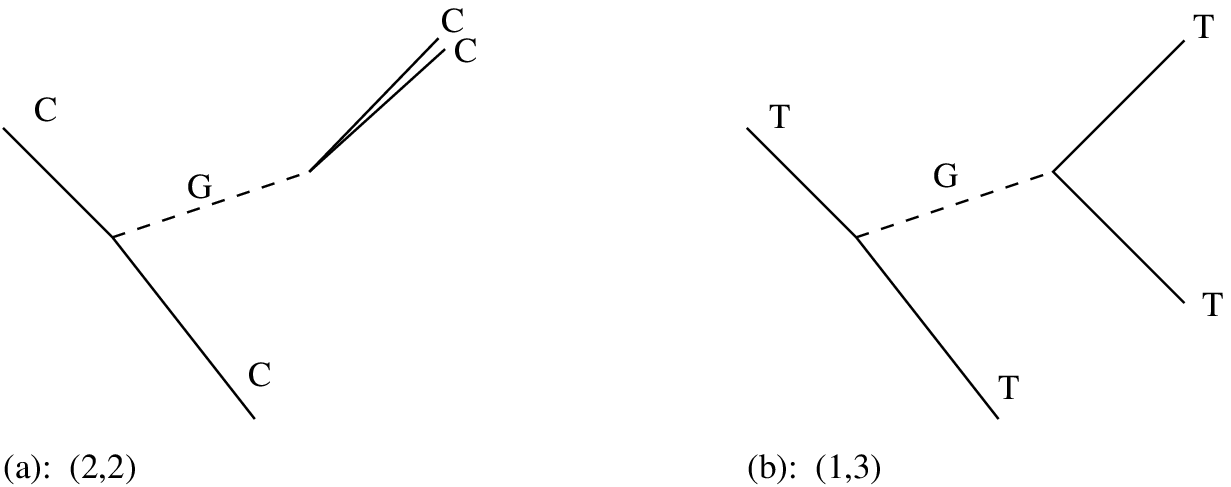}{0.0truein}
Vertex operators involve a term of the form $e^{ \beta\phi - i
\omega t}$. The mass shell condition is\foot{We are still in units
with $\alpha' = \half$.} \eqn\masssh{ (\beta + 2)^2 + \omega^2 = 4
n } where $n$ is an integer in the NS-NS sector and an even
integer in the RR sector. The tachyon and RR axion have $n=0$, the
graviton has $n=1$, etc. Massless particles have $\beta = -2 \pm i
\omega$, where   $+(-)$ correspond to  right (left)-moving
particles. Note that left-moving particles are going into the
strong coupling region, while right-moving particles are going
away from the strong coupling region. Once we choose the
chiralities of the particles we can analytically continue in
$\omega$. But we cannot analytically continue from one chirality
to the other. Another important feature of the bulk amplitudes is
that they preserve left-moving worldsheet fermion number
$(-1)^{f_l}$. The tachyon and right-moving RR fields are odd and
the graviton and left-moving RR field are even.

We are interested in a $(2,2)$ bulk amplitude that contains
two left-moving and two right-moving particles. It was argued in
\Francesco\ that the $(n_l,n_r)$ amplitude (with $n_l$ left movers
and $n_r$ right movers) vanishes if $n_r>1$ and $n_l>1$. Let us
see how this happens in more detail. A direct evaluation of the
$(2 , 2)$ bulk amplitude of four RR particles gives
\eqn\amplitstr{ {\cal A}_{bulk} = \delta(\sum \omega_i) \delta (
\sum \beta_i  + 4) \omega_1 \omega_2 \omega_3 \omega_4 {
\Gamma(t/2) \Gamma(u/2) \Gamma(\half + s/2)
 \over \Gamma(1-t/2) \Gamma(1- u/2) \Gamma(\half - s/2) }
} where we have defined \eqn\defs{\eqalign{ s = & = - {1 \over 4}
[ (\omega_3 + \omega_4)^2 + (\beta_3 + \beta_4 + 2)^2 ] \cr t =
&{\rm same~ with~} 34 \to 31 ~,~~~~~~~~~~u = {\rm same~with~} 34
\to 32  }} The $\delta$ functions in \amplitstr\ enforce
energy-momentum "conservation". The on-shell conditions for the RR
axion \eqn\ons{\omega_i^2+(\beta_i+2)^2=0} then imply
\eqn\stu{s+t+u=1.}

If we take particles 1 and 2 to be left-moving and particles 3 and
4 to be right-moving, then we have
\eqn\dia{\beta_{1,2}=-2-i\omega_{1,2},~~~~~~~~~\beta_{3,4}=-2+i\omega_{3,4}.}
Using the on-shell conditions \ons\ one then finds \eqn\specialk{
s =1 } This in turn  implies that \amplitstr\ is zero.

We wish to understand the cancellations which lead to this result
from the perspective of field theory diagrams. This can be done by
relaxing the on-shell conditions \ons, and defining the off shell
amplitudes by analytic continuation in $s$, $t$ and $u$ in
\amplitstr. It is convenient to do this in such a way that \stu\
is maintained. Lets first look at the $t$ channel. For $s\neq 1$
there are poles at $t = -2n$, $n\geq 0$. These poles correspond to
intermediate NS-NS states with $(-1)^{f_l} =-1$. The first pole
corresponds to the tachyon and it has the simple form
\eqn\simplef{{\cal A}_{t\sim 0} \sim
 { \omega_1 \omega_2 \omega_3 \omega_4 \over t }.
 }
 This agrees with the answer we expect from the effective field theory and
 the coupling \ssi .
 Note that \simplef\ does {\it not} vanish when we set $s=1$, but all higher
 poles do vanish when we set $s=1$. However,
 there is a similar pole in the $u$ channel, which has the same form as
 \simplef\ with $t \to u$. Then \specialk\ implies that
 $u =-t$ so that the two diagrams containing tachyon
 exchange vanish. This also happens in the effective field theory since we have
identical kinematic restrictions.

A different situation arises if we look at the poles in the $s$
channel. Notice that the expression \amplitstr\ has poles at $s =
-1 -2n$. These poles correspond to NS-NS states with $(-1)^{f_l}
=1$. The first state is the graviton. Note that the tachyon does
not get exchanged since the three point coupling vanishes due to
the conservation of worldsheet left fermion number\foot{ In the
effective field theory the couplings coming from \ssi\ vanish if
the two RR axions have the same chirality.}. In other words, the
coupling between two left-moving RR axions and a tachyon vanishes.
The pole at $s= -1$ gives a contribution of the form
\eqn\polegrav{ {\cal A}_{s \sim -1} \sim { \omega_1 \omega_2
\omega_3 \omega_4 \over s + 1 } } as in figure 3a. Note that when
we set \specialk , \polegrav\ gives the contribution due to
graviton exchange which can be interpreted as particle production.
Since in the end the $(2,2)$ amplitude vanishes, this contribution
ends up canceling after including the exchange of all higher
massive states. One might imagine matching \amplitstr\ to a
low-energy action which includes only the tachyon RR axion and
graviton-dilaton. In this case we would simply include the poles
due to \simplef\ and \polegrav\ , then we could think about the
rest as the quartic interaction of the previous subsection (the
$J$ term in \tdi) with a coefficient  simply given by minus
\polegrav\ with $s=1$.

There are other diagrams that could lead to particles on ${\cal I}^+$.
 For example one of the incoming (left moving)  $C $ particles can scatter
into a left moving tachyon and a right moving $C $ 
particle that goes to ${\cal I}^+$. This
process can be explicitly subtracted off if we are interested in
isolating the piece that comes from gravitational 
particle creation. In fact, we
will see that the matrix model computation contains a piece that
comes from this process.

Note that in \NatsuumeSP\ a rather similar diagram, involving an intermediate
graviton, was computed. They sent in a pulse and then they send in a second
pulse and they studied the scattering of the second pulse from the
gravitational field created by the first. In this case the corresponding
bulk amplitude is depicted in figure 3 (b). This is involves a $(3,1)$ bulk
amplitude which does not vanish \dgik \Francesco.
The authors of \NatsuumeSP\ found a detailed agreement between
the effective action computations and the matrix model results, including
a non vanishing $(1,3)$ amplitude.
Note that the graviton which is exchanged in the case studied
in \NatsuumeSP\ can
be on shell. In other words the corresponding invariant is not fixed, but
is a function of the external momenta.

It seems rather surprising that the effects of
{\it static} gravitational fields are seen (\NatsuumeSP)
but time dependent ones are delicately canceled. This might be a consequence
of the $W_{\infty}$ symmetry, and it might be ultimately be the reason that black
holes do not form in the singlet sector \foot{ See \ref\KlebanovVP{
I.~R.~Klebanov,
``Ward identities in two-dimensional string theory,''
Mod.\ Phys.\ Lett.\ A {\bf 7}, 723 (1992)
[arXiv:hep-th/9201005].
}
\ref\KutasovQX{
D.~Kutasov, E.~J.~Martinec and N.~Seiberg,
``Ground rings and their modules in 2-D gravity with c $\leq$ 1 matter,''
Phys.\ Lett.\ B {\bf 276}, 437 (1992)
[arXiv:hep-th/9111048].
} \ref\DasBA{
S.~R.~Das, A.~Dhar, G.~Mandal and S.~R.~Wadia,
``W infinity ward identities and correlation functions in the C = 1 matrix
model,''
Mod.\ Phys.\ Lett.\ A {\bf 7}, 937 (1992)
[Erratum-ibid.\ A {\bf 7}, 2245 (1992)]
[arXiv:hep-th/9112052].
} \ref\DanielssonJF{
U.~H.~Danielsson,
``Symmetries and special states in two-dimensional string theory,''
Nucl.\ Phys.\ B {\bf 380}, 83 (1992)
[arXiv:hep-th/9112061].
}
for further discussion on the implications of the $W_\infty$ symmetry. }.

\newsec{Particle Production in the Matrix Model}

In the previous section, we have computed particle production
arising from various terms in the action.  These results should be
compared with the matrix model computation.  The matrix model has
a nontrivial S-matrix, which is further modified by the leg-pole
factors and then interpreted as the worldsheet S-matrix of string
theory.  In principle, the various terms in the effective action
which we have studied in the previous section can be computed from
this information.  An example of such computation was presented in
\NatsuumeSP, where the coefficients of the Einstein-dilaton term
and the cubic tachyon self interaction where computed.

In the matrix model, particle production at the level at which we
are studying it is related to  the  $1\rightarrow 3$ S-matrix element. This
should not be confused with the statement we made above, which was
that the particle production came from a $(2,2)$ {\it bulk}
amplitude. These two statements are consistent since one of the
particles involved in the bulk computation then scatters from the
wall and comes back out and becomes a right-moving particle also.
The matrix model takes the full process into account. The
computation we will eventually do is really an inclusive
computation where we compute $1 \rightarrow 2 + X$ where $X$ is anything
that comes out late and the $2$ represent the two right-moving
particles that are created at early times. This represents an
incoming particle causing the production of a
particle-antiparticle pair on ${\cal I}^+$. The leading term in
string coupling perturbation theory coming from the matrix model
should reproduce the sum of the various terms computed in the
previous section.

In \NatsuumeSP, the computation starts out as a perturbation
expansion in ${1 \over \mu}$ or equivalently the value of the
string coupling at the tachyon wall.  At the end however, we are
interested in the $\mu$-independent terms, which are a bit tricky
to extract from a ${1 \over \mu}$ expansion. In the bosonic string
considered in \NatsuumeSP, one does not really have the option of
simply working at $\mu=0$ because the theory is ill-defined there.
In contrast for the 0B theory there is no  problem at
$\mu=0$, so we will compute in the relatively simple free-fermion
formalism at $\mu=0$.

\subsec{Basic formulae} In this subsection we will summarize the
basic formulae needed for the computation.

First let us summarize the leg-pole factors.
\eqn\legpolesum{\eqalign{
 \a_{out}(\omega) =& l(\omega)^{-1} \bar \a_{out}(\omega)
 \cr
 \a_{in}(\omega) = & l(\omega)  \bar \a_{in}(\omega)
\cr \l(\omega)_{bos} = & { \Gamma( i \omega) \over \Gamma( - i
\omega) } \cr \l(\omega)_{T}  = & {\Gamma( i \omega/2) \over
\Gamma( - i \omega/2)} \cr \l(\omega)_{C } = & { \Gamma( \half + i
\omega/2) \over \Gamma(\half - i \omega/2)} }} where $\a(\omega)$
is a creation operator for $\omega >0$, overbars denote matrix
model fields, $T$ denotes dressed 0B tachyon and  $C $ denotes 0B
axion. We have chosen $\alpha'=1$ for the bosonic string and
$\alpha' = \half$ for the superstring. The 0B matrix model tachyon
and RR axion are related to the free fermions  by
\eqn\tachsea{\eqalign{
 \psi^R =& e^{ i (\bar T + \bar C )} ~,~~~~~~~~~~~~~~\psi^L =
 e^{ i (\bar T-\bar C  )}
 \cr
 \psi^e = & \psi^L + \psi^R ~,~~~~~~~~~~~~~~~~~\psi^o = \psi^R - \psi^L
 \cr
 \partial \bar T = & \tilde \psi^e \psi^e + \tilde \psi^o \psi^o ~,~~~~~~~~~~~~~~
 \partial \bar C  =   \tilde \psi^e \psi^o + \tilde \psi^o \psi^e
\cr \psi^e \sim &  e^{ i \bar T } (e^{i \bar C  } + e^{- i \bar C
} ) ~,~~~~~~~~~ \psi^o \sim   e^{ i \bar T } (e^{i \bar C } - e^{-
i \bar C  } ) }} $\psi^L$ ($\psi^R$) is localized in the left
(right) part of the Fermi sea, and we use $\tilde \psi$ to
denote $\psi^\dagger$ to avoid a cluttering of indices.
Now we consider the matrix model reflection amplitudes for even and odd
fermions: $\psi^{e}$, $\psi^o$. The reflection amplitudes of
fermions of energy $\epsilon$ are
\eqn\reflfer{\eqalign{
\psi^e_{in}(\epsilon) = &
i e^{-\pi i / 4} 2^{-i\epsilon}
{ \Gamma( {1 \over 4} - i{ \epsilon \over 2} ) \over
\Gamma({1 \over 4} + i {\epsilon \over 2} ) }  \psi^e_{out}(\epsilon)
\cr
\psi^o_{in}(\epsilon) = &
i e^{\pi i / 4} 2^{-i\epsilon}
{ \Gamma( {3 \over 4} - i {\epsilon \over 2} ) \over
\Gamma({3 \over 4} + i {\epsilon \over 2}) }  \psi^o_{out}(\epsilon)
\cr
\psi^{bos}_{in}(\epsilon) = & \sqrt{ \Gamma( \half - i \epsilon)
\over \Gamma(\half + i \epsilon)} \psi^{bos}_{out}
}}
where
$\psi(\epsilon)$ is a fermion creation operator if $\epsilon >0$.
We have also indicated the reflection factor in the bosonic case
(this will make sense only for appropriate Fermi levels).

\subsec{Single fermion reflection}

Let us consider an incoming fermion with a
wavefunction $f_{in}(t+x)$, such that $f_{in}(t) =0$ for $t<0$
\eqn\inwave{ f_{in}(t + x) = {1 \over 2
\pi}\int d\omega f_{in}(\omega) e^{ - i \omega(t+x) } }
Then
$f_{in}(\omega)$ is analytic in the upper half plane. The out
wavefunction is \eqn\outwavefn{ f_{out}(t-x) = \int d\omega
R(\omega) f_{in}(\omega)e^{ - i \omega(t-x)} ~, ~~~~R(\omega) = {
\Gamma( \kappa - i \omega/2) \over \Gamma( \kappa + i \omega/2)} }
An important property of the fermion reflection factor is that it
is analytic in the upper half plane. Naively this would imply that
$f_{out}(t-x)$ is zero for $ t-x <0$. This is not true because $ R
\to \infty$ when $ \omega \to i \infty $. If fact, let us consider
the case where we have a  Fermi level given by $\mu$ and we
consider energies near the Fermi level. Then we just replace
$\omega \to \mu + \omega$ in the reflection factor \outwavefn .
For large $|\mu|$ this reflection factor has a leading phase going
like $R \sim e^{ - i \omega \log|\mu| }$. This means that a low
energy excitation, with energy $\omega \ll |\mu|$ will lead to
\eqn\outfun{ f_{out}(t-x) \sim f_{in}( t - x  + \log |\mu| ) } So
this looks as if the reflection is happening at $ 2 x = \log |\mu|
$. This agrees with the intuition that the reflection occurs at $-
\lambda^2/2 = \mu $, where $\lambda \sim e^{x}$. This is a nice
check of the sign in the argument of the gamma functions \reflfer .

In other words, the early time reflection that we get in the
matrix model is due to the growth of $R$ in the upper half plane
and {\it not} from poles in $R$. On the other hand the late time
behavior of the amplitude does indeed depend on the poles in $R$.

\subsec{One point functions}

We now consider an incoming  coherent state
  \eqn\reft{ | f \rangle = e^{i\int d\o \bar
f(\omega)\bar\ai(\o)}|0\rangle .} where $\bar f^*(\omega)=\bar
f(-\o)$ is the pulse profile in the matrix model and $\bar \ai$ is
a collective field oscillator.\foot{ These obey
\eqn\iil{\bar\alpha(\o)=i\o\sqrt{2}\int ds e^{-i\o
s}\eta(s),~~~~~~
[\bar\alpha(\o),\bar\alpha(\o')]=2\pi\o'\delta(\o+\o').}}
We want to compute \eqn\expcv{
 \langle f | \partial \phi(t) | f \rangle \equiv \langle  \partial \phi(t)
 \rangle_f
 }
Since we are interested in early time behavior on $I^+$, we will
simplify things a bit by taking $f$ not to have an early time tail
and also obey $\int due^{-u}f(u)=0$. In this case the matrix model
$\bar f$ does not contain the leading exponential piece at early
times.

We can compute \expcv\ by expressing  the boson in terms of the
matrix model fermions. For the bosonic string we obtain the
expression  \eqn\expec{\eqalign{
 \langle  \partial \phi(t) \rangle_f \sim & i\int d\omega d \epsilon
e^{- i \omega t}  d u d u' e^{ i ( \omega/2 + \epsilon) u  + i (
\omega/2 - \epsilon )u'}
 \cr
 & l(\omega) R^{-1}( -\omega/2 + \epsilon) R( \omega/2 + \epsilon)
 { e^{ i (\bar f(u) - \bar f(u') )} - 1 \over u - u '}
 }}
Where $R(\omega)$ are the fermion reflection factors \reflfer\ and
the symbol $\sim$ here an hereafter means that factors of 2, -1
and $\pi$ are  omitted. The subtraction of $-1$ corresponds to
normal ordering the fermions. We are interested in the early time
tail, so we want to consider $ t \ll 0$. In this case we may
deform the $\omega$ integral to the upper half plane. Since the
reflection factors are analytic in the upper half plane we see
that (if we do not shift contours in $\epsilon$) we do not get
poles from the bounce factors\foot{ Note that in this case it is
OK to shift the contour to $\omega \to i \infty$ since the growth
of the bounce factors is canceled by the decrease of the leg
factor (away from the purely imaginary axis).}. We do get poles
from the leg factor. For the bosonic string we would get poles at
$ \omega = i n$, $n>0$, and the leading contribution comes from the first
one. So we evaluate the residue of the integrand at $\omega = i$.
 In the bosonic string case the
bounce factors give something of the form \eqn\bouncef{ R^{-1} R
\sim  |\epsilon| \to |\mu| - \epsilon } where the second result is
the one with nonzero $\mu$, $\mu <0$. We now set $u = u_c + r/2$,
$u' = u_c - r/2$. The integrand contains a factor $e^{  i \epsilon
r}$, so the factor of $\epsilon$ may be converted into $ i
\partial_r $, yielding
\eqn\exprf{ -i( |\mu| - i \partial_r )   { e^{ i \bar f(u_c + r/2)
- i \bar f(u_c - r/2)} -1 \over r} |_{r=0} =  ( |\mu| \bar f' +
\half (\bar f')^2 ) } In conclusion, the early time piece is
proportional to \eqn\propearly{ \langle \partial \phi \rangle_f
\sim e^{t} \int du e^u (  |\mu| \bar f' - \half (\bar f')^2 ) }
where we have set $u_c = u$. This answer is valid to all orders in
perturbation theory.
 This expression is
convergent if we choose  $f$ so that the first tail of $\bar f$
vanishes. In other words, we want $\bar f \ll e^{- u}$ for $u \to
\infty$. In general, if we start with a completely localized $f$,
then we get $\bar f \sim e^{ - n u}$ from the leg poles.
A nice derivation of this result was given
in \polchcons , by using the fact that the early time tails are given
by conserved $W_\infty$ charges in the matrix model.
Finally note that \propearly\ can be interpreted as arising from a bulk
interaction of the form $\int e^{- 2 \Phi} T^3 $.

We now compute similar expressions for the 0B case. Let us start
with the tachyon.
 Then we find
\eqn\expec{\eqalign{
 \langle \partial T(t) \rangle_f =& \int d\omega d \epsilon
e^{ -i \omega t}  d u d u' e^{ i ( \omega/2 + \epsilon) u + i (
\omega/2 - \epsilon )u'} l_T(\omega) \times
 \cr
 &  \left[
 R_e^{-1}( -\omega/2 + \epsilon) R_e( \omega/2 + \epsilon)
  \langle : \tilde \psi^e(u') \psi^e(u) : \rangle_f
  \right.
\cr
 & \left. +
 R_o^{-1}( -\omega/2 + \epsilon) R_o( \omega/2 + \epsilon)
\langle : \tilde \psi^o(u') \psi^o(u) : \rangle_f
 \right] }}
As before, we are interested in the upper half $\omega$ plane.
Only the poles of the leg factor contribute. The first is at
$\omega =  2 i$. Then the bounce factors give \eqn\bouncefg{
R_e^{-1} R_e = R_o^{-1} R_o= {1\over 4} ( { 1\over 4} +
\epsilon^2) = {1\over 4} ( { 1\over 4} - \partial_r^2)
~,~~~~~~~~~~ } The first equality in \bouncefg\ holds at all poles
of the leg factor, namely it holds for $\omega = i 2 n $, $n>0$
and $n$ integer. Since the two bounce factor ratios are the same
we get the combination \eqn\comfin{\eqalign{
 \langle : \tilde \psi^e(u') \psi^e(u) :  +
:\tilde \psi^o(u') \psi^o(u):\rangle_f & =
 \langle : \tilde \psi^L(u') \psi^L(u) :  + :\tilde \psi^R(u')
\psi^R(u):\rangle_f \cr & = i{ e^{ i (\bar T(u) - \bar T(u' ) )}
\cos(\bar C (u) - \bar C (u') ) -1 \over u - u' } }} In this and
similar expressions below, $\bar T$ and $\bar C $ to denote their
coherent state expectation values in the coherent state $|\bar
f\rangle $ rather than quantum operators. Hopefully the
distinction will be clear form the context.  The final result at
$\mu=0$ is \eqn\fintypezb{ \langle
\partial T \rangle \sim e^{ 2 t} \int du e^{ - 2 u} [ \bar T' + 4
( \bar C ')^2 \bar T' + { 4 \over 3} ( \bar T')^3
 - { 1 \over 3}  \bar T''' ]
}
Let us comment briefly on this result. The term involving $T^3$ can arise
from a bulk $ e^{-2 \Phi } T^4$ interaction. Notice that the difference
between the bosonic string result and the superstring result comes from the
interplay of the matrix model amplitudes and the leg factors. If we choose a
different values of $\alpha'$ in the bosonic and in the superstring so that
 the matrix
model amplitudes are essentially the same, then the leg pole factors are different.
This difference in the leg-pole factors then translate into the rather different
effective actions in the bosonic string versus the superstring. For example the
bosonic string contains a $T^3$ interaction which is responsible for \propearly ,
while no such interaction is present in the superstring. Indeed, the superstring
result \fintypezb\ is related to $T^4$ interactions in the bulk.

 If we wanted
to consider the result at non-zero $\mu$, then we must shift
$\epsilon$ in \bouncefg\ by $\mu $.
Then we get terms that are at most quadratic in $\mu$.

Now let us consider the same computation for the RR axion
\eqn\expecchi{\eqalign{
 \langle  \partial C (t) \rangle_f =& \int d\omega d \epsilon
e^{ -i \omega t}  d u d u' e^{ i ( \omega/2 + \epsilon) u + i (
\omega/2 - \epsilon )u'} l_C (\omega) \times
 \cr
 &  \left[
 R_o^{-1}( -\omega/2 + \epsilon) R_e( \omega/2 + \epsilon)
 \langle : \tilde \psi^o(u') \psi^e(u) : \rangle
 \right.
\cr
 & \left.+
 R_e^{-1}( -\omega/2 + \epsilon) R_o( \omega/2 + \epsilon)
   \langle : \tilde \psi^e(u') \psi^o(u) : \rangle_f \right] }} As before, we are interested in the upper half $\omega$
plane, and only the poles of the leg-pole factor contribute. The
first one is at $\omega =  i$. The bounce factors give
\eqn\bouncefg{ R_o^{-1} R_e =  R_e^{-1} R_o = {\epsilon \over 2}}
The first equality holds at all poles of the leg factor, namely at
$\omega = i ( 1 + 2 n)$. (Note that we get basically the same as
in the bosonic string \bouncef ) We see that the independent
factors then combine into \eqn\combinci{\eqalign{
 \langle : \tilde \psi^e(u') \psi^o(u) : + : \tilde \psi^e(u') \psi^o(u) : \rangle_f =
& \langle : \tilde \psi^R(u') \psi^R(u) : - : \tilde \psi^L(u')
\psi^L(u) : \rangle_f = \cr & i{ e^{ i (\bar T(u)- \bar T(u'))}
\sin( \bar C (u) - \bar C (u') ) \over u - u'} }} Transforming
\bouncefg\ into a derivative we find that \eqn\finres{ \langle
\partial C  \rangle_f \sim e^t \int du e^{-u} \partial \bar T
\partial \bar C  } This is consistent with a bulk coupling of
the form $T (\nabla C )^2$ (see \ssi ).

\subsec{Axion two point functions}

We now compute the connected part of the axion two point function
\eqn\twopointf{\eqalign{
 ~ \langle \partial C (t) \partial C (t) \rangle_{f,c} =&
\cr & \int d \omega_1 d \omega_1 d\epsilon_1 d \epsilon_2
 du_1 du_2 du'_1 d u'_2
 \cr
  ~&  e^{ - i (\omega_1 + \omega_2)t}
 e^{ i \omega_1(u_1 + u'_1)/2 +i  \epsilon_1( u_1- u_1') }
e^{ i \omega_2(u_2 + u'_2)/2 +i  \epsilon_2( u_2- u_2') }\cr &
 l_C (\omega_1) l_C (\omega_2)
\langle : \tilde \psi_{out}^R(u_1') \psi_{out}^R(u_1) -  \tilde
\psi_{out}^L(u_1') \psi_{out}^L(u_1):
 \cr & : \tilde \psi_{out}^R(u_2') \psi_{out}^R(u_2) -  \tilde \psi_{out}^L(u_2')
 \psi_{out}^L(u_2):\rangle_{f,c} \cr }}
 The $\psi_{out}$ operators in this expression need to be replaced by in operators using the
 bounce factors as above. However, we are eventually going to be interested in in setting
$\omega_1 = i $ and $\omega_2 = i$. So in order not to get a
terrible mess, we are going to evaluate just this term in the full
answer. As we saw above taking  $\omega_1=\omega_2=i$ will imply
that the bounce factors reduce to simple terms involving $ i
\epsilon_1 $ and $i\epsilon_2$. This, in turn become derivatives
with respect to $r_1 = u_1 - u_1'$ and $r_2 = u_2 - u_2'$. So we
get an expression of the form \eqn\twopointfc{\eqalign{
 ~  \langle \partial C (t) \partial C (t) \rangle_{f,c} & = e^{ 2 t} \int
 du_1^c du^c_2 e^{ - u_1^c - u_2^c }\cr
 &~~~~~~  \partial_{r_1} \partial_{r_2}  \langle : \tilde \psi^R(u_1')
 \psi^R(u_1) -  \tilde \psi^L(u_1') \psi^L(u_1):\cr &~~~~~~
 : \tilde \psi^R(u_2') \psi^R(u_2) -  \tilde \psi^L(u_2') \psi^L(u_2):
\rangle_{f,c}|_{r_i =0} \cr }} After taking derivatives,  this
expectation value is proportional to
\eqn\expval{
 { 1 \over  (u^c_1 - u^c_2)^2 } \partial
 \bar C (u_1^c) \partial \bar C (u_2^c)
,}where for simplicity we have taken an antisymmetric incoming
pulse  with $\bar T=0$.

The final answer is \eqn\finexp{ \langle:
\partial C (t) \partial C (t) : \rangle_f \sim e^{2 t} \int du_1
du_2 e^{ - u_1 - u_2} { 1 \over (u_1-u_2)^2 }\partial \bar C (u_1)
\partial \bar C (u_2)
} In treating the double pole we note that it arises from the
contraction
\eqn\contrac{ \psi^*(u_1) \psi(u_2) - :
\psi^*(u_1) \psi(u_2):
 \sim {i \over u_1 - u_2 + i \epsilon }  \sim \psi(u_1)
\psi^*(u_2) - :\psi(u_1) \psi^*(u_2) : } Since the denominator in
\finexp\ came from performing these contractions we should replace
it by $1/(u_1-u_2 + i \epsilon)^2$. It now easy to see that there
is no divergence.

We see that \finexp\ does not have the interpretation of the
Hawking radiation that we were expecting. Nevertheless we would
like to identify the spacetime origin of \finexp . In the
spacetime action we have a coupling of the form \ssi\ \eqn\intver{
S_{int} \sim \int e^{ \Phi} T (\partial C )^2 } This is the same
coupling that was responsible for the one point function \finres .
This coupling can lead to a process where one incoming RR axion
becomes an ingoing  tachyon and an outgoing RR axion. By summing
over the tachyon final states, then we can get \finexp . In fact
the whole answer is accounted for in this fashion. This implies
that all other processes cancel. In fact, this is consistent with
the fact that the $(2,2)$ bulk scattering string amplitude is
zero.

\subsec{Tachyon two point function}

The situation is rather different for the two point function of
the 0B tachyon on ${\cal I^+}$. For the axion, the exponential
tail was produced by the leg pole transform. The falloff rate had
the right strength to correspond to Hawking radiation but, as we
saw, the detailed functional form indicated an alternate source.
For the tachyon, the leg pole transform falls of at twice the
rate, so that  \eqn\ppk{\langle \p T(t) \p T(t)\rangle_f \sim
e^{4t}.} Hence there is no hope of getting anything large enough
to correspond to the rising tail of Hawking radiation.  As
mentioned before, this is not a contradiction because the leading
tail of Hawking radiation can be canceled by other interactions
if they have certain finely-tuned coefficients.
In fact the bulk scattering amplitude to 2 RR fields into 2 tachyons
or 2 tachyons into two 2 tachyons vanishes \Francesco .

\newsec{ Scattering amplitudes for very energetic states}

In this section we consider a single incoming tachyon in the type 0A
theory with very
large energy $\epsilon \gg 1$. For simplicity we set $\mu =0$. We
are then interested in seeing if there is some sign of black hole
formation. Specifically we will compute the number of
particles with energies $\omega$ that we have in the final state.
This computation is very  similar  to those in
\MoorePlesser\
\Martinec . In particular, the analysis of \Martinec\ is more detailed
than what we are going to perform here. We decided to include this because
is it is a rather simple and direct computation.

The exact fermion scattering amplitude is given by\foot{ Up to
phases that are constant or linear in the energy. Such terms do no
matter for this computation.} \eqn\fersca{ R(\epsilon) = {\Gamma(
{1 \over 2} - i {\epsilon \over 2}  ) \over \Gamma( {1\over 2} + i
{\epsilon \over 2} )} } An incoming boson with energy $\epsilon$
may be written as \eqn\incoming{ |\Psi\rangle =
 \alpha_{in}( -\epsilon) |0\rangle =
 \int_0^\epsilon
 dx \psi^\dagger_{in}(x) \psi_{in}(\epsilon -x)  | 0 \rangle
}In terms of out operators
 \eqn\outgoing{
 |\Psi \rangle \equiv \int_0^\epsilon dx  R^*(-x) R(\epsilon -x)
\psi^\dagger_{out}(  x) \psi_{out} (\epsilon -x) |0\rangle } We
can compute the norm of \outgoing\ and we find that it is equal to
\eqn\normp{ \langle \Psi | \Psi \rangle = \epsilon \delta(0) }
where the $\delta(0)$ is just the usual volume factor due to the fact that
we are considering plane waves.
Now we want to compute the expectation value for the number of
particles with fixed energy $ \omega$ in the outgoing state
\outgoing . Namely we are interested in computing \eqn\intercomp{
 \omega N_\omega=\langle \Psi | \alpha_{out}(-\omega) \alpha_{out}(\omega) 
|\Psi \rangle~.}
 Note that leg pole factors cancel out in this computation.  In
order to obtain $N_\omega$ we first compute that
\eqn\computefirst{\eqalign{
\alpha_{out}(\omega) |\Psi \rangle = & \int_{\omega}^\epsilon dx F(x)
\psi_{out}^\dagger(x-\omega) \psi_{out}(\epsilon - x) |0 \rangle
\cr F(x) =& R(-x)^* R(\epsilon - x) - R^*(-(x- \omega))
R(\epsilon - x + \omega) }}
 where we assumed that $\omega
\not = \epsilon$ (otherwise there is an additional term). Note
$\alpha_0$ annihilates the state \incoming, since  $\alpha_0$ is
the fermion number operator and the net fermion number of
\incoming\ is zero.

Now we go back to \intercomp\ which is equal to the norm of the
state \computefirst . We find that \intercomp\ becomes (up to a
$\delta(0)$) \eqn\resultgen{\eqalign{
 & \int_\omega^\epsilon dx
\left| R^*(-x) R(\epsilon -x) - R^*(-(x - \omega)) R(\epsilon - x
+ \omega) \right|^2 = \cr & \int_\omega^\epsilon dx \left| 1 -R(-x)
R^*(\epsilon -x) R^*(-(x - \omega)) R(\epsilon - x + \omega)
\right|^2 = \cr & \int_\omega^\epsilon dx ( 2 - 2 Re(R R^* R^* R) )
} } where we used that $|R|=1$. Using Stirling's formula we
can show that the leading phase at  large $x$ is \eqn\largeal{
{\Gamma( {1\over 2} - i {x\over 2} ) \over \Gamma( { 1 \over 2} +
i {x \over 2}  ) } \sim  e^{ - i { x }  \log  x }
  } We then see that for large $x$
\eqn\resxep{ R(-x) R^*(-(x - \omega)) \sim e^{ i \omega \log x} }
Similarly we obtain that for large $\epsilon -x$
 \eqn\simresf{ R^*(\epsilon - x)
R(\epsilon - x + \omega) \sim e^{ - i  \omega \log (\epsilon
- x) } }
We conclude that the product of all phases is \eqn\productph{ x^{  i
\omega} (\epsilon - x)^{-  i \omega} = t^{  i \omega} (1 -
t)^{- i \omega} ~,~~~~~~~ {\rm where}~~~~ t = { x \over \epsilon}
}

We now substitute \productph\ in the last term of the integral
\resultgen\ and change the integration variable to $t$. We need
to evaluate an integral of the form
\eqn\endup{
 \int^1_{\omega/\epsilon}  dt t^{  i \omega} (1-t)^{-  i \omega} =
 \Gamma( 1 -  i \omega) \Gamma( 1 +  i \omega)  + o(\omega/\epsilon) \sim
  {  \pi \omega
\over \sinh  \pi \omega}~. }
 Putting all this together we find that
\intercomp\ has the form
\eqn\expectval{ \langle \omega N_\omega \rangle = { \langle \Psi | 
\alpha_{out}(-\omega) \alpha_{out}(\omega) 
| \Psi \rangle \over  \langle \Psi | \Psi
\rangle } = 2 ( 1 - {  \pi \omega \over \sinh  \pi \omega} )  ~,~~~~~~~~{\rm for}~~
\omega \ll \epsilon }
Notice that this result goes to
zero for $\omega \to 0$ as expected.

The expectation value \expectval\ goes to a
constant for large $\omega$. If the physical process were  black
hole formation we would have expected a thermal factor with a
temperature of order one. We might imagine that the first, constant, factor
represents a reflected pulse that arises at the moment we form a black hole
and that the second factor is proportional to Hawking radiation. Indeed the
second factor has an exponential decay reminiscent of Hawking radiation.
 But the
second factor has the
wrong sign and its integral is independent
of $\epsilon$ \foot{ By writing
 $ { 1 \over \sinh \pi \omega} \sim { e^{ - \pi \omega} } \sim e^{ - \beta \omega}$
  for
large $\omega$ in \expectval\ we can read off  the
a temperature temperature which is the same as the temperature
of  the 0A SL(2)/U(1) black hole, $\beta
= 2 \pi \sqrt{ \alpha'/2} $.  }.
The expectation value
\expectval\ is saying that we have an energy of order one in each
mode of frequency $\omega$ (for $ 1 \ll  \omega \ll  \epsilon$).
Hence there is no indication from this computation that a highly
energetic tachyon forms a black hole. Similar observations were
made in \refs{\MoorePlesser,\Martinec}.

\newsec{Discussion}

We close with a discussion of several issues concerning
black holes in the matrix model in the context of our results.

\subsec{The definition of time}
One issue concerns the time coordinate to be used in the
spacetime/matrix model dictionary. In spacetime computations,
there are several natural         retarded time coordinates on ${\cal
I}^+$.
One of them is where the conformal factor $\rho$ asymptotically
vanishes
and the linear
dilaton is time-independent. We refer to these as inertial coordinates, 
as they are associated to asymptotically inertial observers.
The null coordinates
\eqn\metrga{ ds^2 = - dt^+_{inertial} dt^-_{inertial} +
g_{++} (dt^+_{inertial})^2 } can be seen to be inertial in this sense.

A more common coordinate choice (employed for example in this paper and
\NatsuumeSP )
is conformal gauge
\eqn\ththree{ ds^2=- e^{2 \rho}dt^+_{conformal}dt^-_{conformal}~.}
The conformal retarded time and inertial retarded times
are related at early times on ${\cal I}^+$ via
\eqn\resz{t^-_{conformal}=t^-_{inertial}-\half
U(\infty)e^{2t^-_{inertial} }+\cdots}
These two coordinates systems define outgoing vacua on ${\cal I}^+$ which
are related by a nontrivial Bogolubov transformation. If we work in
the conformal gauge, Hawking particle production is seen indirectly as a
consequence of this Bogolubov transformation to inertial coordinates.
In the inertial coordinates \metrga,
particle production arises directly from the coupling of
the time-dependent metric component $g_{++}$ to the massless field $S$.

In order to compare the matrix model with the spacetime results
we must know whether the matrix model result corresponds to
conformal, inertial, or yet another
coordinate choice.
Although the literature is not very explicit on this point, in most
discussions of the matrix model it is implicitly assumed that the matrix model
time $t_{mm}$ and eigenvalue are related to the spacetime retarded time
via
\eqn\res{t_{mm} - \ln \lambda =t^-_{inertial}.}
This assumption has been adopted in this paper in equation \ttg, 
and we have seen herein that the full string theory answer is consistent
with the matrix model answer only when we make this (standard) assumption.


In this paper we are considering normalizable excitations of the
ground state. In the context of large, non-normalizable perturbations, it is
less obvious how time should be defined. 
Some discussion is in \refs{\aj, \finn, \aaj}.


\subsec{The non-singlet sector}

In this paper we have analyzed whether black holes could form in the singlet
sector. Critics have pointed out that our efforts were doomed to fail 
because it is
known that the $SL(2)/U(1)$ black hole involves non singlets. However, it is
also known that the $SL(2)/U(1)$ solution is not the only one that looks like a
black hole\foot{
There are configurations that look like black holes with different radii 
\kkk .}. So it might well be that there are other solutions, which exist
purely in the singlet sector, which also have a black hole interpretation, even
if they are not the standard $SL(2)/U(1)$ black hole.
Our results, however, are consistent with the idea that there are no
black holes in the singlet sector.
So a plausible explanation for the peculiar horizon-aversion
of collapsing shells found in section 3 is that a black hole cannot be
formed because the matter is in the singlet sector and the black hole is
not.


If the non-singlet nature of black holes is the explanation, one
may also ask if black holes can be formed from non-singlet initial data.
For example one might try to form them by collapsing D0-branes,
which also can be argued to involve non-singlets. It would also be
nice to know what the matrix model dual to a  {\it Lorentzian}
black hole is, as opposed to euclidean black hole \kkk .
\subsec{$SL(2)_k/U(1)$=black hole for small $k$?}

For large $k$, Witten's $SL(2)_k/U(1)$ coset is a weakly coupled
sigma model describing a black hole. For small $k$ the sigma model
becomes strongly coupled so it is no longer clear if it should be
thought of as a black hole. In the matrix model context, the
relevant cosets always are in the small $k$ regime. Therefore it
is not clear if they should be interpreted as black holes, as we
discuss in this section.

As it was discussed in \ref\jmstrings{J. Maldacena, talk at strings 2004.} the
nature of the $SL(2)_k/U(1)$ conformal field theory
depends in an important way on $k$. For large $k$, $k > 3$, this conformal
field theory has a normalizable zero mode which corresponds to changing the
value of the dilaton at the tip\foot{ Actually, the theory has two
zero modes one changes the value of the dilaton and the other changes the
integral of $B_{NS}$ on the cigar.}.
For $k\leq 3$ this mode becomes non-normalizable. This can be seen from the
algebraic analysis of the $SL(2)_k/U(1)$ model as follows.
The zero mode is constructed from discrete representations as
$J_{-1}^- \bar J^-_{-1}| j , m \rangle$ with $j=m=1$\foot{
In conventions as in \ref\MaldacenaHW{
J.~M.~Maldacena and H.~Ooguri,
J.\ Math.\ Phys.\  {\bf 42}, 2929 (2001)
[arXiv:hep-th/0001053].
}. See \ref\KutasovXU{
D.~Kutasov and N.~Seiberg,
JHEP {\bf 9904}, 008 (1999)
[arXiv:hep-th/9903219].
} for further discussions about this state.}.
  This is a state
with worldsheet conformal weight one. In general, a state is
normalizable only if $ \half < j < {k \over 2} - \half $,
 so this state with
$j=1$ is normalizable only for $k> 3$. From the target space point
of view this seems surprising. It is clear that the wavefunction
of the vertex operator that changes the  dilaton  at the tip
decays as $e^{ -2 \phi}$ for $\phi \to \infty$ while the
normalizability condition only requires that it decays faster than
$e^{ - \phi}$ \foot{ The last statement is saying how we are
normalizing $\phi$.}. The crucial point is   that a mode on the
cigar does not have definite winding at infinity. A good way to
think about it is to imagine that the wavefunction of a given mode
contains all possible winding numbers. In particular, the
wavefunction of this mode goes as \eqn\modewf{ V_{\delta \Phi_0}
\sim J_{-1}^- \bar J_{-1}^-e^{- { 2} \phi}  +
 e^{ - 2 {(k/2 -1)} \phi } e^{ + i \sqrt{k} (X_L-X_R) }  +
 e^{ - 2 {(k/2 -1)} \phi } e^{ - i \sqrt{k} (X_L-X_R) } + \cdots
}
where
the dots indicate higher winding modes.
We see that the winding number one
terms in \modewf\ are not normalizable for $k\leq 3$.

The conclusion from these remarks is that the cigar conformal
field theory at $k\leq 3$ does not have a zero mode and
furthermore, when we go to the weakly coupled region, it differs
from the linear dilaton theory by a non-normalizable operator.
This operator is the one appearing in the second term in \modewf .
Note that we have derived all this from analysis purely in the
$SL(2,R)_k/U(1)$ theory. It is not necessary to assume that there
is a dual sine-Liouville theory. Of course, the fact that the
operator is not-normalizable is clear in the sine-Liouville theory
and in a different context it was noted in \ref\KutasovUA{
D.~Kutasov and N.~Seiberg, ``Noncritical Superstrings,'' Phys.\
Lett.\ B {\bf 251}, 67 (1990).
}. Even though this can be derived purely from the
$SL(2,R)_k/U(1)$ CFT, these facts cannot be derived from the
lowest order in $\alpha'$ analysis of the sigma model, even if we
use the supposedly ``exact'' metrics in the literature.
 Any analysis that focuses
only on the lowest order in $\alpha'$ will miss the winding modes,
which are not included. The point is that the curvatures are large
and it is necessary to solve the theory exactly in $\alpha'$.

For $k>3$ the existence of this zero mode
is conceptually related to the fact that the asymptotic value of the radius
of the cigar cannot be changed, any attempt to change the value of the radius
will drive  the value of the dilaton at the tip to zero or infinity. On the
other hand, for $k<3$ there is no zero mode, so we expect that it should be
possible to change the value of the radius. Indeed, the matrix model analysis
of \kkk\ shows that the value of the radius can be changed.

All of these points are highlighting the fact that the Lorentzian continuation
of the cigar conformal field theory is not obvious. In particular, we 
would need to
understand the Lorentzian interpretation of the extra winding mode operator that
is being turned on\foot{A naive Lorentzian's continuation of the matrix model in
\kkk\ is not possible because it would lead to a Lagrangian that is not local
in time, since the Wilson lines are in the exponent. Maybe a local Hamiltonian
might be obtained after ``integrating in" some quark degrees of freedom...}.

One can, nevertheless, be very naive and compute the absorption
cross section for the black hole. If the absorption is nonzero it
is reasonable to call the geometry a black hole. For this one can
compute the momentum two point function in the cigar geometry and
analytically continue the result to Lorentzian signature. This
gives a non-zero answer for the absorption  \ref\GiveonWN{
A.~Giveon, A.~Konechny, A.~Pakman and A.~Sever,
``Type 0 strings in a 2-d black hole,''
JHEP {\bf 0310}, 025 (2003)
[arXiv:hep-th/0309056].
}\dvv , even when $k\le
3$. If we believe this continuation it can be taken as evidence
that the $SL(2,R)_k/U(1)$ CFT is a black hole even for $k\le 3$.

\subsec{Conclusions}

Previous investigations of high energy tachyon 
scattering \MoorePlesser \Martinec ,
have shown that it does not lead to the formation of black holes. 
In those studies scattering of a single high energy tachyon was 
considered. In this paper we considered
the collapse of a  coherent state formed by many tachyons. By looking at the
time at which the energy is emitted we concluded that there cannot be a long
lived black hole. The energy is emitted within a time of order $\log E$ of the
the time at which the black hole would be formed. 
If a black hole had formed we would
have expected that a finite fraction of the incoming energy would come out
over a time proportional to $E$. This simple argument does not rule out
the formation of a black hole with very high temperature.

On a different, but related note, we have also asked 
here the following question:
to what extent is the low energy effective action corrected? We know from the
analysis of \NatsuumeSP , that many of its features are indeed consistent 
with the matrix model. In particular, \NatsuumeSP\ have shown that 
the matrix model includes
gravitational interactions which are present in the linear dilaton theory,
in the region far from the Liouville wall. Particle creation is another effect
mediated by the gravitational field, so our naive expectation was that
we were going to see it if we repeated the analysis of \NatsuumeSP .
However, here there is an interesting cancellation between the
gravitational effect and the effects coming from the exchange of all massive
fields. This
cancellation, which is present in the exact string theory answer, implies
that there is no net particle creation by an infalling pulse.

Our computation of particle creation by the infalling pulse focused just
on the leading term. It would be nice to see what happens at higher orders in the
$e^t$ expansion.

\centerline{\bf Acknowledgments}

We are grateful to I. Klebanov, F. Larsen, J. Polchinski, S. Shenker, N.
Seiberg, T. Takayanagi and H. Verlinde for useful discussions.
This work was supported in part by DOE grant DE-FG02-91ER40654 and
the Harvard Society of Fellows.

\listrefs
  \end